\documentclass[prb,epsf,epsfig]{revtex4}
\usepackage{amsmath}
\usepackage[toc,page]{appendix}
\usepackage{graphicx,pdflscape}
\usepackage{hyperref}

\usepackage{epsfig}
\usepackage{graphicx}
\usepackage{dcolumn}
\usepackage{bm}
\usepackage{color}

\newcommand{\dispeq}[1]{Eq.~(\ref{#1})}

\newcommand{\tJ}{\ $t$-$J$ \ }

\newcommand{\beq}{\begin{equation}}
\newcommand{\eeq}{\end{equation}}
\newcommand{\barray}{\begin{eqnarray}}
\newcommand{\earray}{\end{eqnarray}}

\newcommand{\nn}{\nonumber}

\newcommand{\disp}[1]{Eq.~(\ref{#1})}

\newcommand{\refdisp}[1]{Ref.~(\onlinecite{#1})}
\newcommand{\figdisp}[1]{Fig.~(\ref{#1})}

\DeclareMathOperator{\re}{Re}
\DeclareMathOperator{\sgn}{sgn}

\begin{document}

\title{ Renormalization Group Study of a  Fragile Fermi liquid in $1+\epsilon$ dimensions   }
\author{ Peizhi Mai$^\dagger$, H. R. Krishnamurthy$^{*\dagger}$ and B. Sriram Shastry$^\dagger$}
\affiliation{$^\dagger$Physics Department, University of California,  Santa Cruz, Ca 95064 }
\affiliation{$^*$Physics Department, Indian Institute of Science, Bangalore 560012, INDIA}
\date{\today}
\begin{abstract}
We present a calculation of the low energy Greens function of interacting fermions in $1+\epsilon$ dimensions using the method of extended poor man's scaling,   developed here. We compute the wave function renormalization $Z(\omega)$ and also the decay rate near the Fermi energy.  Despite the lack of  $\omega^2$ damping characteristic of  3-dimensional Fermi liquids, we show that  quasiparticles do  exist in $1+\epsilon$ dimensions,  in the sense that the quasiparticle weight $Z$ is finite and that the damping rate is smaller than the energy. We explicitly compute the crossover from this behavior to a 1-dimensional type Tomonaga-Luttinger liquid behavior at higher energies.

 \end{abstract}
\maketitle


\section{Introduction}
 Recent experimental work ~\cite{E0,E1,E2,E3,E4,E5} on the angle resolved photoemission (ARPES) have investigated weakly two dimensional systems. These are  equivalently viewed as   weakly coupled 1-dimensional  chains, and  exhibit the characteristics of 1-dimensional Tomonaga-Luttinger type systems with anomalous dimensions, exhibiting a  crossover at lowest energies to a Fermi liquid type behavior, with a finite but very small value of  the quasiparticle weight $Z$. The small scale of $Z$ here is related to the almost 1-d nature of the systems. Fermi liquids with a small but non zero $Z$ also arise in other important condensed matter systems in 0, 2 , 3 and $\infty$ dimensions. A small $Z$ in  the latter   arise due to strong correlations, rather than reduced dimensionality. 
   Historically the Gutzwiller wave function \cite{Gutzwiller} provided a  first example of  such a behavior,   suggesting   a strong correlation induced  vanishing $Z$ near the Mott insulating state. This was made  especially explicit in the work of Brinkman and Rice \cite{Brinkman-Rice}.  In 0 dimensions,  the asymmetric single  Anderson impurity model \cite{Hewson-rg,RG1,RG2,RG3,RG4} (AIM) provides a well studied and classic example.  Here one finds an exponentially small $Z\sim e^{-\frac{1}{2 (1-n_d)}}$ from the Bethe Ansatz solution \cite{Hewson-Rasul},  in the limit where the occupancy of the impurity level $n_d \to 1$. In the $d=\infty$ Hubbard model, which is solvable numerically by the dynamical mean field theory \cite{dmft} (DMFT), one finds a vanishing $Z\sim (1-n)$ as the electron density tends to the Mott insulating value $n \to 1$, with possibly small corrections\cite{ecfl-dmft} to the exponent for very small $(1-n)$. In other dimensions various approximations- such as the slave particle field theories- suggest a similar small  value  of $Z$ in the metallic state found near the Mott insulating limit.    We may provisionally call this group of metallic systems with a small $Z$,  whatever the origin of its small scale, as ``Fragile Fermi Liquids'' (FFL).

 We next consider the important issue of the damping rate in order to refine this notion. Recent work on the large U Hubbard or the \tJ model using extremely correlated Fermi liquid (ECFL) theory \cite{ecfl-dmft, Gweon-Shastry, Shastry-Asymmetry, ecfl, 
ecfl-aim} gives an interesting insight into the nature of the quasiparticle damping  near the insulating limit, which agrees in remarkable detail with the results of DMFT\cite{ecfl-dmft}. In the large d model at low energies, one finds that the quasiparticle Greens function at the Fermi momentum, including the  damping, can be expressed as
\beq
\left[ G^{-1}(k_F, \omega+ i 0^+) \right]_{\omega\to 0} \sim  \frac{\omega}{Z}  + i \, \frac{1}{\Omega_c} \,  \left( \frac{\omega}{Z} \right)^2   - i  \frac{1}{\Omega^2_d} \,  \left( \frac{\omega}{Z} \right)^3 + \ldots ,  \label{damping-g}
\eeq
where $\Omega_c$ and $\Omega_d$ are energies on the scale of the bandwidth. Since 
the imaginary part gives us the damping of the quasiparticles  here,  this expression goes beyond the domain of the Landau Fermi liquid theory.  The Landau theory  merely says that the damping is of $O(\omega)^2$
without giving the scale of the damping, nor does it specify the terms beyond the leading order. Thus the Greens function including damping exhibits an $\omega/Z$ scaling, with an unexpected and prominent odd in $\omega$ corrections to damping as in \disp{damping-g}. 
This cubic term helps  in understanding the ARPES line shapes in {\em very strongly correlated metals} as shown in~\refdisp{Gweon-Shastry,Shastry-Asymmetry}, and also in the thermopower of correlated matter \refdisp{Palsson-Kotliar}. It  may be viewed as one of the  signatures of extreme correlations, in addition to their role in diminishing  $Z$.   For the AIM,  a similar expression for the low energy Greens function to quadratic order results in the extension of the Fermi liquid theory in the interesting work of Hewson\cite{Hewson-rg}. 
In the following we  focus  on effects of small $Z$ brought about by dimensionality rather than strong correlations. Therefore we shall be content to ignore  the cubic term and discuss the leading quadratic term alone.  
 Taking the above examples as benchmarks, we refine the notion of the Fragile Fermi Liquids.
 These  may be characterized as having  quasiparticles endowed  with a small $Z$,  with a  damping (smaller than the energy) on an energy scale  that   itself shrinks with $Z$.

In order to explore further this notion of Fragile Fermi liquids, it would be of value to have solvable models that give detailed results for the damping, along with the required small Z.
 In this work we study  weakly coupled $1+\epsilon$ dimensional systems resulting in a Fermi liquid where $Z$ is very small, as described in the first paragraph.  In view of the physics described by \disp{damping-g}, our goal is to compute not only $Z$, but also the damping of the quasiparticles, through a controlled calculation within a  $1+\epsilon$ dimensional model system, with $\epsilon >0$.  We expect that the quadratic behavior of the damping in \disp{damping-g} would be lost in the case of $1+\epsilon$ dimensions, but nevertheless the damping would be small relative to the energy of the quasiparticle. It is of interest then to check if the $\omega/Z$ scaling survives, to the extent possible with the proximity of the Tomonaga-Luttinger behavior at exactly 1-d. For this purpose we study a sufficiently simple model that allows an asymptotically exact calculation, using the renormalization group, of the 
low energy Greens function, including the damping. This would also enable us to study the crossover from a Fragile Fermi liquid at the lowest energies, to a Tomonaga-Luttinger type behavior at higher energies, and thereby make contact with the experiments ~\cite{E0,E1,E2,E3,E4,E5}. The  model considered is the simplest one in $1+\epsilon$ dimensions, and is essentially the same as the one studied in the early work of Ueda and Rice (UR) in~\refdisp{T1}. For our purposes, it turns out to be necessary to compute the scale (or frequency) dependent $Z(\omega)$, and not just the static limit of this object. We will denote the static limit as $Z(0)\to Z$. Furthermore, we are able to calculate the crossover from high to low energy behavior on a crossover scale that depends on $\epsilon$. At low energies we obtain asymptotically a Fragile Fermi Liquid behavior:  the leading damping term of \disp{damping-g} becomes $ i \, \frac{1}{\Omega_c(\omega)} \,  \left( \frac{\omega}{Z} \right)^2 $ with an $\omega$ dependent energy scale $\Omega_c(\omega)$. The result is summarized as:
\barray
Z  & =  &  \exp\{- d_0\frac{{\eta}^{3/4}}{\sqrt{\epsilon}}\}, \label{result-Z} \\
\frac{2}{\Omega_c(\omega)} &=&  \frac{d Z(\omega)}{d \omega}, \label{result-damping-1}
\earray
where $\eta$ is the anomalous dimension in 1-d and $d_0$ is a constant around $1.09$, 
with a singular low energy behavior of 
\beq Z'(\omega) \sim Z(\omega) \times \frac{1}{ 2^\epsilon (1+ \epsilon)} \frac{|\omega|^{\epsilon-1}}{(\log|\omega|/2)^2}. \label{result-damping-2}
\eeq
In view of the singularity of $Z'(\omega)$ the final behavior of the damping term at small $\omega$ is
$\sim |\omega|^{1+\epsilon}/{\log(|\omega|/2)^2}$, which is smaller than  the energy of the particle  $ |\omega|$.  Putting these together  we  find
\beq
\left[ G^{-1}(k_F, \omega+ i 0^+) \right]_{\omega\to 0} \sim  \frac{\omega}{Z} + \frac{i }{2^{1+\epsilon}(1+\epsilon)}\,  \frac{1}{(\log(|\omega|/2))^2}\,\times\left( \frac{|\omega|}{Z}\right)^{1+\epsilon}  ,
 \label{gsimple3}
\eeq
exhibiting an $\omega/Z$ scaling,  apart from the weak logarithmic correction and  setting $Z^\epsilon\to 1$.

 It is amusing to note that although our calculation  \disp{result-damping-2} is designed for $\epsilon \ll 1$, if   pushed  somewhat bravely to $\epsilon \sim 1$, suggests that the singularity of $Z'(\omega)$  in 2 dimensions would be weak, and give rise to a quadratic damping with possibly $\log \omega$ corrections.  This is indeed correct as we know from other works. Using the full solution of the crossover problem, we compute the spectral function of an electron at the Fermi point from high to lowest energies, for a few typical values of the initial coupling constants.

  We next summarize the literature and discuss what is the new result in this paper. 
 UR performed a renormalization group (RG) analysis for small $\epsilon$ and showed that for $\epsilon >0$ a Fermi liquid (FL) fixed point emerges, while $\epsilon=0$ has a line of fixed points which maps to  the Tomonaga-Luttinger model \cite{Giamarchi} with anomalous dimension $\eta$ (defined below in greater detail). This line of fixed points  arises from the competition between the Peierls and Cooper channels.    Further interesting theoretical  work on this model has been undertaken in \refdisp{T2,T3,T4,T5,T6,T7}. For instance at small $\epsilon$, Castellani, Di Castro and Metzner \refdisp{T4} computed the  value of  the quasiparticle weight $Z$, their result is again a non analytic dependence on $\epsilon$, with a slightly different set of exponents $Z\sim \exp\{- \frac{{\eta}}{{\epsilon}}\}$, and should be compared with  our result for $Z$  reported in \disp{result-Z}.
 For a fixed $\eta$ our expression \disp{result-Z} would give a somewhat bigger magnitude of $Z$, but it is  qualitatively similar.    As far as we are aware \disp{result-damping-1}, \disp{result-damping-2} and \disp{gsimple3} are  new.

Also new in our work is the method we use for our calculations. Our results stated above require the calculation of  the full $\omega$ dependent  self energy $\Im m \Sigma(k\sim k_f,\omega)$ for the model of Ueda and Rice.  For this purpose  we have developed a renormalization group (RG) procedure  that is a modification of the Wilsonian RG approach for fermions,  presented pedagogically in the excellent review by Shankar~\cite{Shankar}. The modification becomes necessary because the approach outlined in Ref. \onlinecite{Shankar} leads to difficulties when one tries to use it for calculating the frequency dependent self energy. The difficulties as well as the main features of the new method, which we refer to as the Extended Poor Man's Scaling (EPMS) prescription because it is very much in the spirit of Anderson's celebrated ``poor man's scaling'' approach to the Kondo problem\cite{RG1}, are discussed briefly in the next section, and in greater detail in Section \ref{EPMS}. Our calculations also employ a different simplification of the momentum integrations that arise in $1+ \epsilon$ dimensions as compared to the discussion in Ref. \onlinecite{T1}. The new simplification is also summarized in the next section, and presented in detail in in Section \ref{EPMS}. We emphasize that this simplification is merely for ease of calculation, and we expect that the physics of significance that we discuss will be valid beyond the simplification.

\section{The Model}
\begin{figure}
  \centerline{\epsfig{file=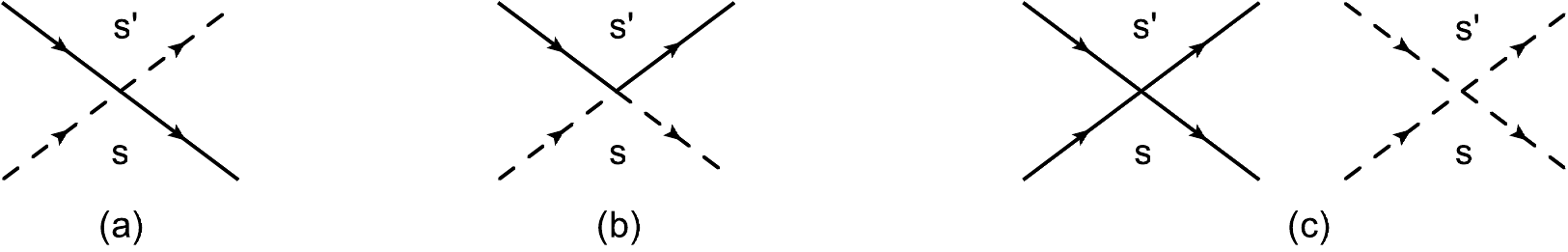,width=0.6\columnwidth}}
  \caption{\label{g1g2g4} We use solid line for fermions at the left branch and dashed line for fermions at the right branch. (a) $g_1$ represents backward inter branch scattering; (b) $g_2$ represents forward inter branch scattering; (c) $g_4$ represents intra branch scattering.   }
\end{figure}

The partition function~\cite{Schulz,Solyom} for the model of interacting fermions  in one dimension (1-d) without umklapp processes (and assuming zero temperature) that we study in this paper can be written as the Fermionic functional integral
\begin{eqnarray}
{\cal Z}=\int [{\cal D}\phi]\textrm{e}^{S(\phi)},
\end{eqnarray}
\barray
S(\phi)=  &&\sum_{s,\alpha=L,R}\int^\infty_{-\infty} \frac{d\omega}{2\pi}\int^{\varLambda_0}_{-\varLambda_0}\frac{dk}{2\pi} (i\omega-\varepsilon_\alpha(k)) \phi^{*}_{s\alpha}(k\thinspace\omega)\phi_{s\alpha}(k\thinspace\omega)
+\sum_{s,s'}\int_{k\thinspace\omega;\varLambda_0} V_{int}[\{\phi\}]  \nn\\
V_{int}[\{\phi\}]=&& g_1 \phi^{*}_{s,L}(1)\phi^{*}_{s',R}(2) \phi_{s',L}(3) \phi_{s,R}(4)\Delta
+ g_2\phi^{*}_{s,L}(1)\phi^{*}_{s',R}(2)\phi_{s',R}(3) \phi_{s,L}(4)\, \Delta 
\nn \\
&&+\sum_{\alpha=L,R}\frac{g_4}{2}\phi^{*}_{s,\alpha}(1) \phi^{*}_{s',\alpha}(2)\phi_{s',\alpha}(3) \phi_{s,\alpha}(4)\, \Delta ,\label{1da}
\earray
where $\phi$ and $\phi^*$ are Grassman numbers; $\alpha=R$ and $\varepsilon_R(k)=k$ for the right branch; $\alpha=L$ and $\varepsilon_L(k)=-k$ for the left branch; $\omega$, $\varLambda_0$ and $k$ are dimensionless quantities defined respectively as $\omega=\omega_{ph}/(\varLambda_{0ph}v_F)$, $\varLambda_0=\varLambda_{0ph}/\varLambda_{0ph}=1$, $k=(k_{ph}+k_F)/\varLambda_{0ph}$ in the left branch and $k=(k_{ph}-k_F)/\varLambda_{0ph}$ in the right branch with $\omega_{ph}$, $k_{ph}$, $\varLambda_{0ph}$ being the \textit{physical} Matsubara frequency,  momentum and  momentum cutoff respectively. Furthermore, we have used the abbreviated notations:  $\prod_{j}\int^\infty_{-\infty} \frac{d\omega_j}{2\pi}\int^{\varLambda_0}_{-\varLambda_0}\frac{dk_j}{2\pi}\to \int_{k\thinspace\omega;\varLambda_0}$, $\phi^{*}_{s,\alpha}(k_j\thinspace\omega_j)\to\phi^{*}_{s,\alpha}(j)$, $\phi_{s,\alpha}(k_j\thinspace\omega_j)\to\phi_{s,\alpha}(j)$ and $\delta(k_1+k_2-k_3-k_4)
\delta(\omega_1+\omega_2-\omega_3-\omega_4)\to\Delta$. The dimensionless coupling constants $g_j=g'_j/v_F$, where $g'_{j}$ have their usual meanings as the coupling constants used in the literature, some times referred to as ``g-ology"~\cite{Solyom}, with $g'_{1}$ or $g_1$  corresponding to the backward (Fig.\ref{g1g2g4}(a),  i.e., $(k_F,-k_F)\to (-k_F,k_F))$, and $g'_2$ or $g_{2}$  to the forward (Fig.~\ref{g1g2g4}(b), i.e., $(k_F,-k_F)\to (k_F,- k_F))$ \textit{inter}-branch scattering terms,  and $g'_4$ or $g_{4}$ corresponding to the \textit{intra}-branch forward scattering term (Fig.~\ref{g1g2g4}(c)).

As is well known\cite{Solyom}, standard diagrammatic perturbation theory in powers of the coupling constants $g_1, g_2$ and $g_4$ for the self energy and other  properties of the model in \disp{1da} in 1-d lead to (logarithmic) divergences. Such divergences are best handled by scaling or RG approaches\cite{Solyom,Shankar,T1} either of which leads to the same scaling equations for the effective coupling constants as a function of a "running" momentum cutoff as high momentum fermion degrees of freedom are recursively eliminated and the cutoff is continuously reduced. As mentioned in the introduction, in this work we are interested in performing a detailed RG calculation of the frequency dependent self energy and Greens function of the model in \disp{1da}, which is clearly more demanding than finding the scaling equations for the  coupling constants. Furthermore, we would like to extend such a calculation to $1+ \epsilon$ dimensions where we should be able to see the Fermi liquid emerging from a non Fermi liquid state.

In extending the existing  Wilsonian RG prescription, which is well explained in Shankar's review article, to the calculation of self energy in  $1+\epsilon$ dimensions, we  encounter two difficulties. The first difficulty is that the Wilsonian  RG,  becomes cumbersome for performing calculations of self energies and in particular the quasiparticle weights~\cite{Shankar,Metzner}. The reason is that the rules of Wilsonian RG require that {\textit all} the momentum labels in the internal propagators in diagrammatic perturbation theory, equivalent to intermediate excited states in traditional perturbation theory, must correspond to the fast or high momentum degrees of freedom that are being eliminated.  As discussed in detail in Section III, momentum and energy conservation then leave the self energy unchanged until the running momentum cutoff is half of the original momentum cutoff, and this renders the method difficult to implement.
To solve this problem, we propose a modification of Wilson's scheme,  whose mode elimination process is  in the spirit of Anderson's poor man's scaling approach~\cite{RG1} to the Kondo problem.  We will refer to as the Extended Poor Man's Scaling (EPMS).
It differs from the Wilson scheme in  that it only requires the intermediate states that are eliminated to involve  \textit{at least one} high energy or fast mode, while in the Wilson scheme  \textit{all} the eliminated states are required to involve only fast modes. This procedure leads to contributions to the self energy arising continuously from the very beginning of the reduction of the momentum cutoff, and makes it easier to track its evolution from high to low frequency scales. The procedure is argued to be self-consistent for the current problem, and as a check we verify that the various exponents and other properties calculated using the new procedure agree with available results from the literature.

The second difficulty has to do with the angular integrals that arise in extending the calculations to non-integer dimensions. To deal with this, we propose a simpler prescription for dealing with $1+\epsilon$ dimensions than used earlier\cite{T1}, which we argue is valid when $\epsilon$ and $\omega$ are small (see Section \ref{EPMS}.D for the details). Using such a prescription and the  second order EPMS method, we obtain the flow equations for the coupling constants and for the $Z$ factor, and numerical as well as exact limiting results for the $Z$ factor and $\Im m \Sigma$. They all show crossover behaviors, with the emergent crossover  scale being given by  $l^*=1/\epsilon$ or $\omega^*=2\textrm{e}^{-1/\epsilon}$, where  $l=\ln(\varLambda_0/\varLambda)$ ($\varLambda$ is the running cutoff). When $l\ll l^*$ or $\omega\gg\omega^*$, the system shows 1-d-Tomonaga-Luttinger type  behavior, while it approaches the higher dimensional limit and shows Fermi liquid behavior if $l\gg l^*$ or $\omega\ll\omega^*$. Also, we show that when $\epsilon<\eta$ , where  $\eta$ is the anomalous dimension from the 1-d limit,  one obtains a ``Fragile Fermi Liquid'' low energy behavior, with extremely small $Z$.

The rest of this paper is organized as follows. In Section \ref{EPMS}, we discuss the difficulties in calculating the self energy and the Z factor using Wilsonian RG in slightly greater detail, and then outline the EPMS method and our prescription for calculations in $1 + \epsilon$ dimensions. In Section \ref{g1g2}, we discuss the second order flow equations for the coupling constants and their solutions obtained using our prescriptions and show that they are in agreement with those in the literature. Sections \ref{zz} and \ref{sigmaomega} contain our central, new results for the $Z$ and for the leading behavior of the self energy in  $1+\epsilon$ dimensions. Section \ref{symmetrybreaking} has a brief discussion about the breaking down in $1+\epsilon$ dimensions of the `Lorentz-invariance' that is a characteristic feature of the asymptotic (low $\omega,k$) behavior of correlation functions of interacting fermions in 1-d. In Section \ref{conclusion} we summarize the main points of the paper. Since we make repeated comparisons of the EPMS method to the Wilsonian RG, for convenience we have summarized the salient aspects of the latter in Appendix \ref{Shankar-RG}. The full details of the EPMS prescription are presented in Appendix \ref{fulldetail}. Readers who are unfamiliar with RG calculations in the context of 1-d fermionic systems are likely to find the rest of the paper more accessible if they go through the Appndix A first.

\section{The Extended Poor Man's Scaling method and The $1+\epsilon$ expansion prescription\label{EPMS}}

\subsection{Difficulties in calculating the Self Energy and the Z factor using Wilsonian RG}

\begin{figure}
  \centerline{\epsfig{file=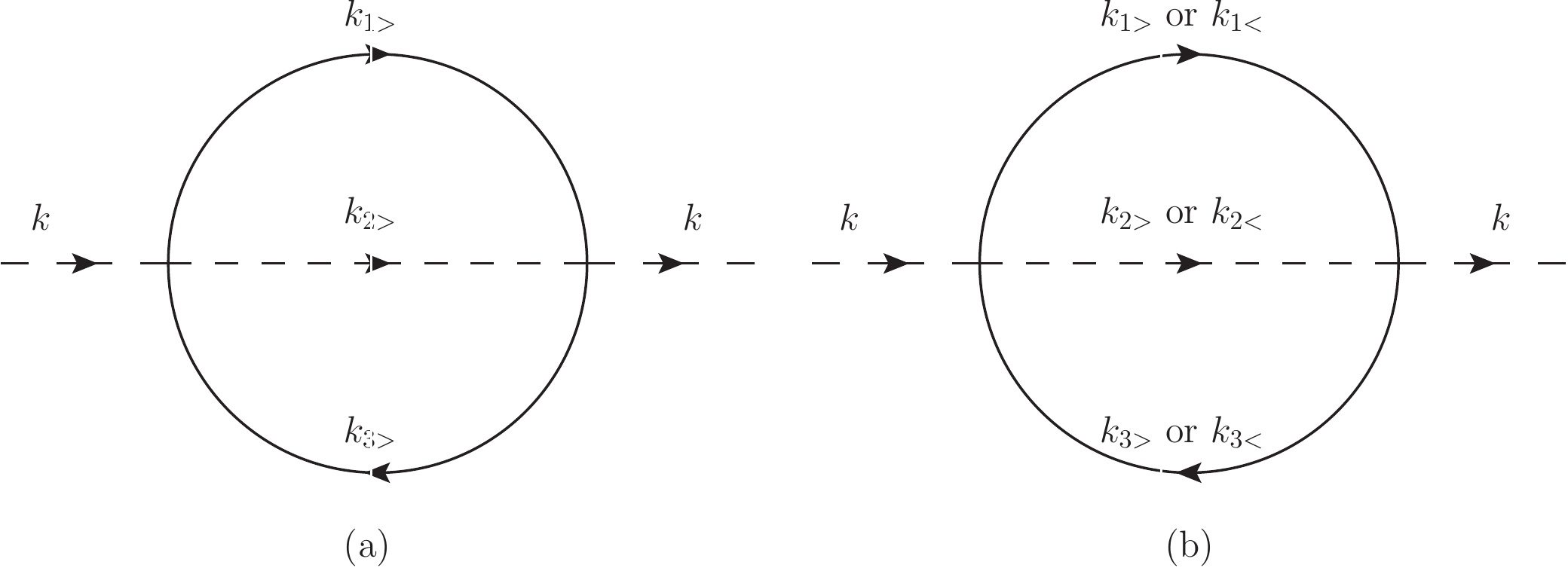,width=0.5\columnwidth}}
  \caption{\label{sunrise}(a) In Wilsonian RG, we calculate diagrams with all internal modes as fast modes. (b) In EPMS, we calculate diagrams with at least one internal mode as fast mode.
  }
\end{figure}

In this subsection, we discuss the difficulties in calculating the self energy and the Z factor using Wilsonian RG in slightly greater detail. In particular,  the quasiparticle weight $Z$ comes from the frequency derivative of the self energy, with (the external) $k=0$ .  In Wilsonian RG, the first $\omega$ dependent contribution to the self-energy comes from the two-loop ``sunrise'' diagrams like  the one shown in \figdisp{sunrise} (a). In a one-dimensional system, the  contribution from this diagram to the  self-energy at a certain step of the RG is proportional to the integral (using rescaled internal momenta and frequencies as in \disp{Seff} of the Appendix \ref{Shankar-RG})
\beq
\int_{d\varLambda_0} \frac{dk'_1}{2\pi}\int_{d\varLambda_0} \frac{dk'_2}{2\pi}
\int^\infty_{-\infty}\frac{d\omega'_1}{2\pi}\int^\infty_{-\infty}\frac{d\omega'_2}{2\pi}
\frac{1}{i\omega'_1+k'_1}\frac{1}{i\omega'_2-k'_2}
\frac{\theta_s(k'_1+k'_2-k')}{i(\omega'_1+\omega'_2-\omega')+(k'_1+k'_2-k')}.\label{simplify}
\eeq
where the delta functions arising from frequency and momentum conservation have been used to carry out the integral over $\omega'_{3}$ and  $k'_{3}$. The subscripts   $d\varLambda_0$ on the (momentum) integral signs are used to denote the constraints that the integrated momenta belong to the \textit{eliminated} shell:  $\varLambda_0/s<|k'_{1,2}|<\varLambda_0$, and the function $\theta_s(k'_1+k'_2-k')$, with $\theta_s(x)$ defined to be zero unless $\varLambda_0/s<|x|<\varLambda_0$, keeps track of the same constraint on $k'_1+k'_2-k'$. For $k=0$, $k'=0$. By carrying out the frequency integrals using contour integration, it is straightforward to verify that non-vanishing contributions to the integral  for $k'=0$ can come only from the regions $(k'_1>0$, $k'_2<0$, $k'_1+k'_2<0)$ or $(k'_1<0$, $k'_2>0$, $k'_1+k'_2>0)$. Either of these is incompatible with the momentum shell constraints on $k'_1,k'_2$ \textit{and} $k'_1+k'_2$, unless\cite{Kopietz2} $s>2$. So the above contribution to the self energy vanishes for $1<s<2$, and  there would certainly be no contribution from \textit{infinitesimal} mode elimination, with $s=1+dl$. Thus, nonvanishing frequency dependent contributions to the self energy  can arise  only from the one loop (or Hartree) diagrams involving the \textit{frequency dependent} two body and three body vertexes like the ones in \figdisp{23ver}, which are ``irrelevant" in the RG sense. But even this contribution will not appear in the first few steps of the RG, i.e., not until the running cutoff is reduced to  $\varLambda_0/2$. Thus it becomes cumbersome to calculate self-energy contributions beyond one-loop using the Wilsonian RG~\cite{Metzner}. Therefore we propose the EPMS scheme, which makes the  calculation of two loop contributions to the self energy relatively easier.

\subsection{The EPMS method}


Now we introduce the procedure of EPMS by elucidating the similarities and the differences between EPMS and Wilsonian RG. EPMS is different from Wilsonian RG~\cite{Shankar} in its way of mode elimination. In Wilsonian RG, we calculate  diagrams with the constraint that \textit{all} the internal modes are  only fast modes. In EPMS, we calculate the diagrams with the modified constraint that \textit{at least one} of the internal modes is a fast mode.  In this sense, EPMS can be regarded as a field theory version of Anderson's poor man's scaling method~\cite{RG1}. The sunrise diagrams in \figdisp{sunrise} attempt to depict the difference by way of  an example. We note that EPMS still retains the spirit of RG  in that we integrate out high energy degrees of freedom and study the low energy effective theory~\cite{Shankar,RG2}. As discussed above,  Wilsonian RG is not very convenient for calculating the two loop contributions to the self energy because the non-vanishing contributions come from formally irrelevant two or three body vertexes produced in previous steps, and furthermore do not appear until the running cutoff reduces to $\varLambda_0/2$. EPMS proposes to overcome this difficulty by taking into account all the contributions (that would have arisen in the subsequent steps of Wilsonian RG) from some of the formally irrelevant two and three body vertexes  at the same time when those vertexes are produced. With this idea, non-vanishing contributions to the frequency dependent self energy, for example, appear at the very first step of EPMS, and are accumulated continually  from the EPMS mode elimination process. The same form of effective action is obtained after mode elimination in EPMS  as in \disp{Seff} (see Appendix A) from Wilsonian RG, but with different multiplicative renormalization factors $a$, $b$ and $c$. Apart from this, the steps involving the rescaling of frequencies, momenta, and fields in EPMS are the same as in Wilsonian RG.

In the following, we use second order renormalization of the one body vertex (or self energy) as an example. In Wilsonian RG, when we calculate the sunrise diagram in \figdisp{sunrise}(a) in the $(n+1)^{th}$ step (i.e., when the running cutoff is reduced from $\varLambda_n \equiv \varLambda_0/s^n$ to $\varLambda_{n+1}\equiv \varLambda_0/s^{n+1}$), \textit{all} the (rescaled) internal momenta being integrated out are restricted to the shell $d\varLambda_0$ as in \disp{simplify}. However, in EPMS, while \textit{one} of the internal momenta being integrated out is still restricted to the shell, all the others could be either fast modes or slow modes, as depicted in \figdisp{sunrise}(b). The simplest way of doing this is as follows: First, prior to the $(n+1)^{th}$ step of EPMS, we calculate the net self energy to second order in the current values of the leading coupling constants:

\beq
\begin{split}
I(k,i\omega;\varLambda_n)&=[g(\varLambda_n)]^2\int_{-\varLambda_0}^{\varLambda_0}\frac{dk_{1n}}{2\pi} \int_{-\varLambda_0}^{\varLambda_0}\frac{dk_{2n}}{2\pi}\int_{-\infty}^{+\infty}\frac{d\omega_{1n}}{2\pi} \int_{-\infty}^{+\infty}\frac{d\omega_{2n}}{2\pi}
\\&\frac{1}{i\omega_{1n}+k_{1n}}\frac{1}{i\omega_{2n}-k_{2n}}\frac{1}{i(\omega_{1n}+\omega_{2n}-\omega_n) +(k_{1n}+k_{2n}-k_n)},
\\&=[g(\varLambda_n)]^2s^n\int_{-\varLambda_n}^{\varLambda_n}\frac{dk_1}{2\pi}\int_{-\varLambda_n}^{\varLambda_n} \frac{dk_2}{2\pi}\int_{-\infty}^{+\infty}\frac{d\omega_1}{2\pi}\int_{-\infty}^{+\infty}\frac{d\omega_2}{2\pi}
\\&\frac{1}{i\omega_1+k_1}\frac{1}{i\omega_2-k_2}\frac{1}{i(\omega_1+\omega_2-\omega)+(k_1+k_2-k)}.\label{III}
\end{split}
\eeq

Here, $\omega_{jn}\equiv s^n\omega_j$, $k_{jn}\equiv s^nk_j$ denote the rescaled internal frequencies and momenta, and $\omega_{n}\equiv s^n\omega$, $k_n \equiv s^nk$ the rescaled external frequency and momentum; like in \disp{simplify}, the momentum and frequency conserving delta functions have been used to calculate the integrals over $\omega_{3n}$ and $k_{3n}$,  with the remaining rescaled internal momenta  being \textit{fully} integrated, from $-\varLambda_0$ to $\varLambda_0$; and $[g(\varLambda_{n})]^2$ is the square sum of the running coupling constants discussed and defined in Section \ref{zz}. Then we take the difference

\beq
\Delta I(k, i\omega; \varLambda_n, \varLambda_{n+1})=I(k, i\omega; \varLambda_n)-\frac{I(k, i\omega; \varLambda_{n+1})[g(\varLambda_{n})]^2}{s[g(\varLambda_{n+1})]^2},
\eeq

as the \textit{incremental} contribution to the self energy from the $(n+1)^{th}$ mode elimination step of the EPMS program. The factor $[g(\varLambda_{n})]^2/(s[g(\varLambda_{n+1})]^2)$ is used in order to retain the same running coupling constants and relevance as in $I(k, i\omega; \varLambda_n)$. From this we calculate the \textit{multiplicatively cumulative} contributions to the renormalization coefficients $a$ and $b$  introduced in Appendix A as,

\beq
\tilde{a}_e(\varLambda_m \rightarrow \varLambda_{m+1})=1+\frac{\partial \Delta I(k, i\omega; \varLambda_m, \varLambda_{m+1})}{\partial (i\omega_m)}\Big|_{\omega\rightarrow 0,k\rightarrow 0}\label{a2}
\eeq
and
\beq
\tilde{b}_e(\varLambda_m \rightarrow \varLambda_{m+1})=1+\frac{\partial \Delta I(k, i\omega; \varLambda_m, \varLambda_{m+1})}{\partial k_m}\Big|_{\omega\rightarrow 0,k\rightarrow 0},\label{b2}
\eeq
where the subscript "$e$" is used to denote that the contributions are  from EPMS. 

\subsection{Additional rules of EPMS}
Although EPMS overcomes the difficulties of Wilsonian RG in calculating the self-energy contributions, it has some disadvantages. In Wilsonian RG, there is no divergence in any intermediate step because the upper and lower limits of integration are always finite numbers with the same sign. But there is no guarantee of this  in EPMS; it is certain to work only when divergences that could in principle be present in $ I(k, \omega; \varLambda_n)$ get cancelled in calculating $\Delta I(k, \omega; \varLambda_n, \varLambda_{n+1})$. The logarithmic divergence in one-dimension is an example.

\begin{figure}
  \centerline{\epsfig{file=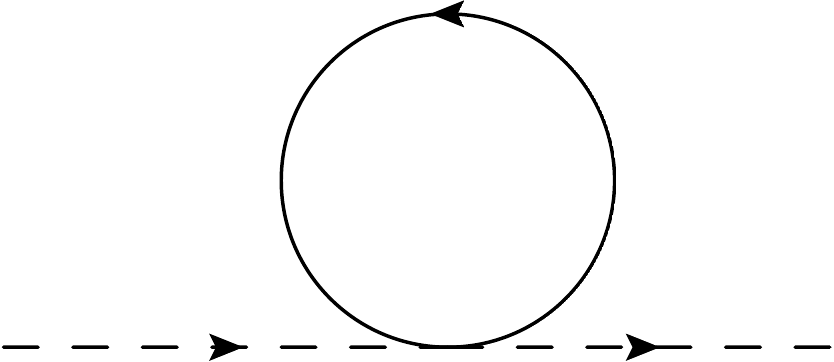,width=0.3\columnwidth}}
  \caption{\label{fog}first order diagram contributing to one body vertex}
\end{figure}

Also, as discussed in the last subsection, the true difference between EPMS and Wilsonian RG is in the order in which diagrams are being summed. What is produced at a certain step in EPMS includes not only the contributions from that very step of Wilsonian RG, but also a set of terms from later steps and of higher order. This poses the problem of avoiding double counting in EPMS. In order to resolve the double counting issue, we have to add some additional rules into the EPMS procedure. First of all, given a specific order of calculations, only the highest order diagrams and tree diagrams are calculated with running coupling constants. For example, \figdisp{sunrise} and \figdisp{23ver}(a) are the highest order loop diagrams in a second order calculation. On the contrary, the lower order diagram like \figdisp{fog} should be calculated with bare(original) coupling constant. Second, the contribution to lower order vertexes from formally irrelevant higher order vertexes is calculated using the original coupling constants. For example, in the Fermi gas model the original couplings of irrelevant vertexes are zero. So there will be no contribution from irrelevant vertexes to lower order vertexes in EPMS.

A more detailed discussion of the comparison between the EPMS calculations and the Wilsonian RG calculations is presented in Appendix \ref{fulldetail}.

\renewcommand{\vec}[1]{\mathbf{#1}}

\subsection{The $1+\epsilon$ expansion prescription\label{onepluse}}

\begin{figure}
  \centerline{\epsfig{file=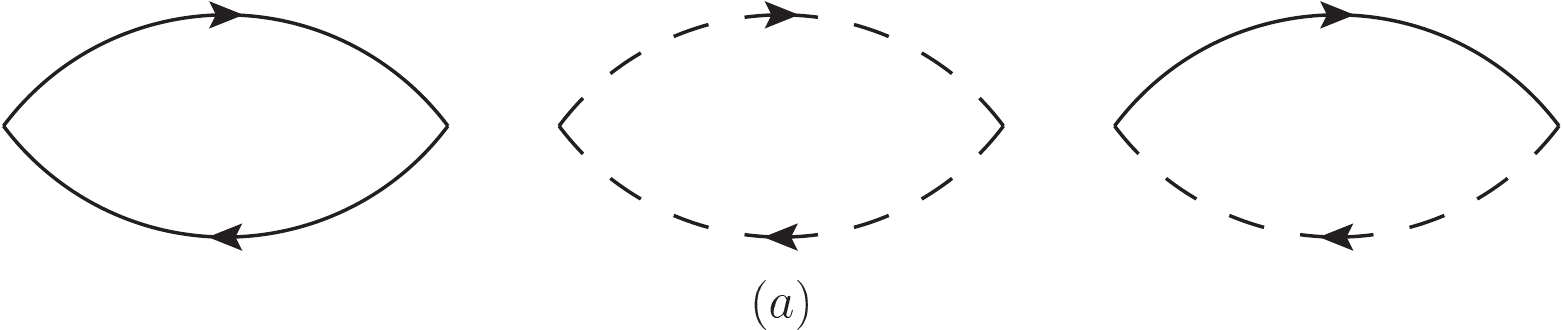,width=0.4\columnwidth}~~~~~~~~~~~~~~\epsfig{file=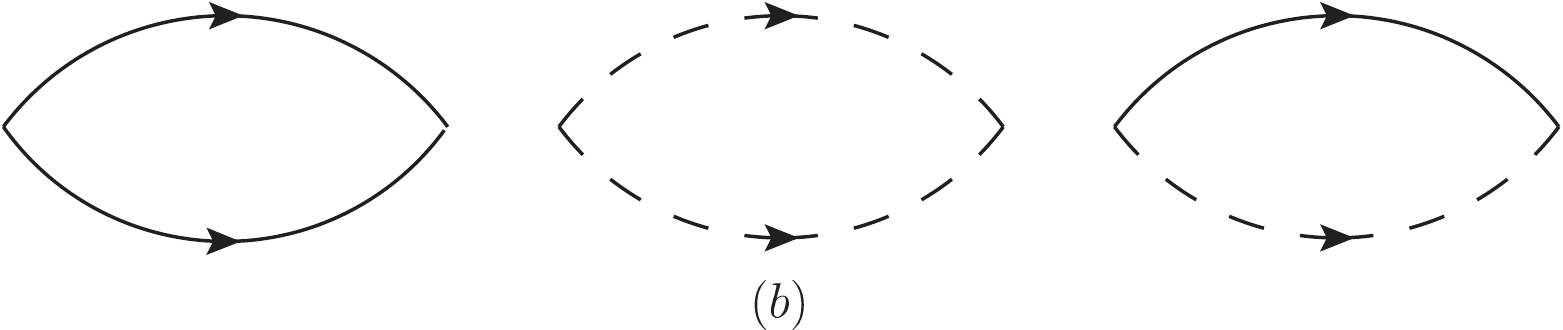,width=0.4\columnwidth}}
  \caption{\label{PC} (a) Peierls Channels; (b) Cooper Channels}
\end{figure}

The angular integrals that arise  when one implements RG calculations in dimensions larger than 1-d are in general rather difficult to evaluate. In this paper, we are particularly interested in $1+\epsilon$ dimensions with $\epsilon \ll 1$. Drawing inspiration from  \refdisp{T1}, we use the following prescription which should be valid for small values of $\epsilon$ and the external frequency $\omega$. It relies on the fact~\cite{T1} that the Cooper (particle-particle) channels (see \figdisp{PC} b) do not depend sensitively on dimensionality while the Peierls (particle-hole) channels (see \figdisp{PC} a) do. This asymmetry can be understood as follows. For the marginal one-dimensional Cooper channels, the momentum transfer is zero, which means the two incoming (or outgoing) momenta are equal and opposite. In dimensions higher than 1,  the outgoing momenta can be at an arbitrary angle relative to the incoming momenta. This property leads to the Cooper (BCS) instability in one, two and three dimensions~\cite{Shankar}. On the other hand, in the marginal case for the one-dimensional Peierls channel, the momentum transfer is $2k_F$. In higher dimensions, the angle between incoming and outgoing momenta is strongly restricted if the momentum transfer is fixed and nonzero. Therefore the Peierls instability is suppressed by the angular integral in higher dimensions.

We propose the following simple prescription for $1+\epsilon$ expansion by considering this asymmetry as the leading effect arising from the extra $\epsilon$ dimensions that needs to be taken into account. Hence, when the Cooper channels are calculated in $1+\epsilon$ dimensions, we use the same formula as in 1-d. However, when calculating the Peierls channels in $1+\epsilon$ dimensions, we introduce an additional factor $|k|^\epsilon$  in an appropriate momentum integral. For example, in the case of sunrise diagram like the ones in \figdisp{sunrise}, the momentum integral over $k_3$ (the one in the opposite direction relative to the other two) should include the factor $|k_3|^\epsilon$.  This introduction of $|k|^\epsilon$ is to be regarded as a purely  mathematical device to approximately take into account the crucial effects of the extra $\epsilon$ dimensions. We note also that in the RG calculation, the $k$ in $|k|^\epsilon$ should always be in terms of the original scale. Otherwise, the rescaling of $k$ in $|k|^\epsilon$ would lead to the changes in the relevance of different terms and get in conflict with the fact that the relevancy of each term is the same in one, two and three dimensions~\cite{Shankar}. We show in the next section that this prescription gives the same flow equations for the coupling constants as in \refdisp{T1}.

In principle, there could be other slightly different  schemes\cite{Metzner} for $1+\epsilon$ dimensions. The reason for choosing our scheme is that it introduces the higher dimension effects without changing the interaction effects in 1-d qualitatively. In a 2-d system, there are three classes of interactions\cite{Metzner,Shankar}, i.e., back, forward and exchange scattering interactions. If we apply this 2-d classification of interactions directly in $1+\epsilon$ dimensions, both $g_1$ and $g_2$ in the 1-d model get regarded as back scattering terms. Then it would be hard to connect to 1-d case as well as look at the crossover behaviors. Instead, our prescription in $1+\epsilon$ dimensions could be imagined as saying that for both the $g_1$ and $g_2$ terms, the incoming as well as outgoing momenta are equal and opposite, but the outgoing momenta could be a bit off the incoming line. And we still take $g_1$ and $g_2$ as back and forward scattering interactions respectively. Such a generalization of the 1-d model does not change the nature of $g_1$ and $g_2$ in 1-d qualitatively and hence helps to understand the crossover behaviors between Tomonaga-Luttinger liquid in 1-d and Fermi liquid in higher dimensions.

\begin{figure}
  \centerline{\epsfig{file=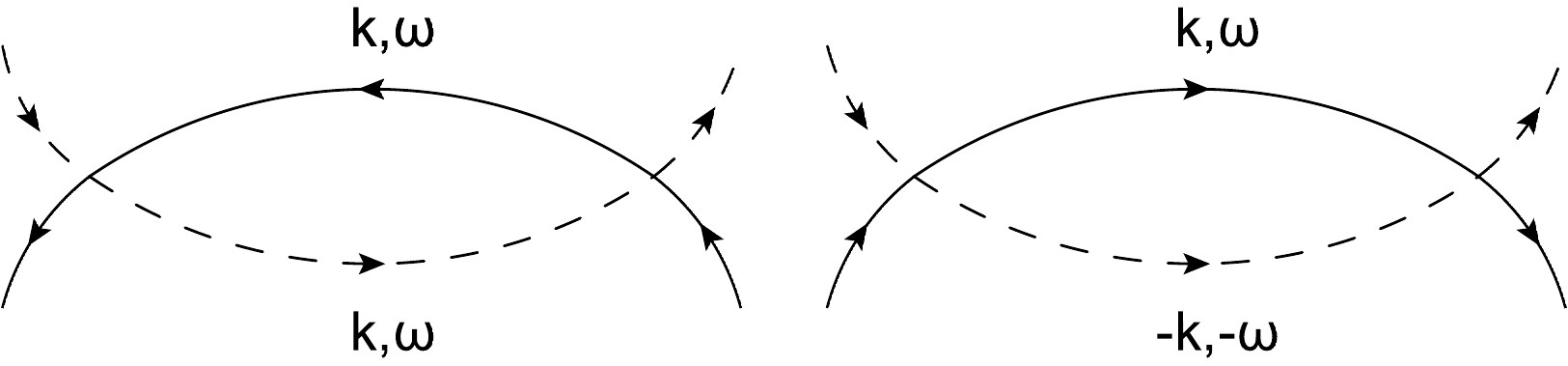,width=0.5\columnwidth}}
  \caption{\label{mtbv}All non-vanishing one loop diagrams contributing to marginal two body vertex
  }
\end{figure}


\section{Second order flow equations for the coupling constants}\label{g1g2}
In this section, we derive the second order flow equations for the coupling constants in 1+$\epsilon$ dimensions. If we only look at the contribution to the marginal couplings, corresponding to all external momenta  being  at the Fermi surface (external $k=0$), the only non-vanishing diagrams are shown in \figdisp{mtbv} (the Peierls and Cooper channels involving the same branches are easily shown to give vanishing contributions) and the internal momenta would \textit{both} have to be high momenta. Therefore, EPMS will give the same results as Wilsonian RG~\cite{Shankar}. Since the two body vertexes are marginal, in calculating the diagrams we can use momenta and frequencies as per the \textit{original} scale and change the limits on the momentum integrals to take into account the  running cutoff $\varLambda$ without changing the result.

The Peierls channel contribution with $k_1,\omega_1$ and $k_1,\omega_1$ on different branches is:
\beq
\pi^0=\int^{\varLambda}_{\varLambda/s}\frac{|k_1|^\epsilon dk_1}{2\pi}\int\frac{d\omega_1}
{2\pi}\frac{1}{i\omega_1+k_1}\frac{1}{i\omega_1-k_1}+\int^{-\varLambda/s}_{-\varLambda}
\frac{|k_1|^\epsilon dk_1}{2\pi}\int\frac{d\omega_1}{2\pi}\frac{1}{i\omega_1+k_1}
\frac{1}{i\omega_1-k_1}=-\frac{\textrm{e}^{-\epsilon l}dl}{2\pi},
\eeq
where the last result is obtained for infinitesimal change in the running cutoff, as obtained by setting $s=\textrm{e}^{dl}$, and $\varLambda=\varLambda_0/s^{n}=\textrm{e}^{-n \, dl}=\textrm{e}^{-l}$. The Cooper channel with $k_1,\omega_1$ and $k_1,-\omega_1$ on different branches gives
\beq
\Delta^0=\int^{\varLambda}_{\varLambda/s}\frac{dk_1}{2\pi}\int\frac{d\omega_1}{2\pi}\frac{1}{i\omega_1-k_1}
\frac{1}{-i\omega_1-k_1}+\int^{-\varLambda/s}_{-\varLambda}\frac{dk_1}{2\pi}\int\frac{d\omega_1}{2\pi}
\frac{1}{i\omega_1-k_1}\frac{1}{-i\omega_1-k_1}=\frac{dl}{2\pi}.
\eeq
Again, the last result above is for infinitesimal RG. 

Then, for the EPMS step reducing the running cutoff from $\varLambda$ to $\varLambda/s$, the incremental change in the coupling constants is given by $\delta g_1=2g_1^2\pi^0-2g_1g_2\Delta^0-2g_1g_2\pi^0$, $\delta g_2=-g_1^2\Delta^0-g_2^2\Delta^0-g_2^2\pi_0$ and $g_4=0$.
Hence we get the flow equations:
\barray
\frac{dg_1}{dl}&=&-\frac{g_1g_2}{\pi}-\frac{(g_1^2-g_1g_2)\textrm{e}^{-\epsilon l}}{\pi},\label{floweqng1}  \\
\frac{dg_2}{dl}&=&-\frac{g_1^2+g_2^2}{2\pi}+\frac{g_2^2\textrm{e}^{-\epsilon l}}{2\pi},\label{floweqng2} \\
\frac{dg_4}{dl}&=&0.
\earray
These equations are essentially the same as in \refdisp{T1}, except that $\epsilon$ appears here instead of $\epsilon/2$ in the reference. The equations show an emergent crossover  scale $l^*=1/\epsilon$. For $l<<l^*$, the $1+\epsilon$ dimensional equations behave like their 1-d versions~\cite{Solyom}:  $dg_1/dl=-g_1^2/\pi$, $dg_2/dl=-g_1^2/(2\pi)$ and $dg_4/dl=0$. The solutions are
\beq
g_1(l)=\frac{g_{10}}{1+g_{10}l/\pi},
\eeq
\beq
g_1(l)-2g_2(l)=g_{10}-2g_{20},
\eeq

\begin{figure}
  \centerline{\epsfig{file=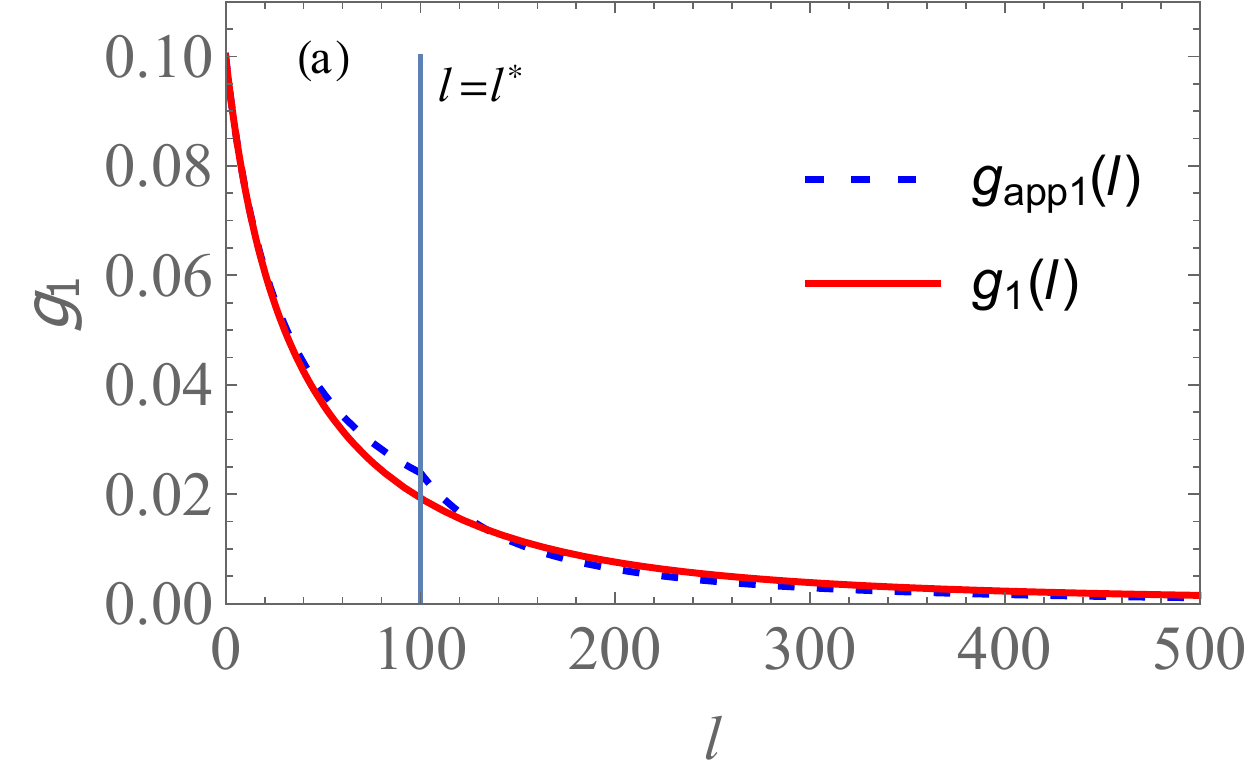,width=0.5\columnwidth}\epsfig{file=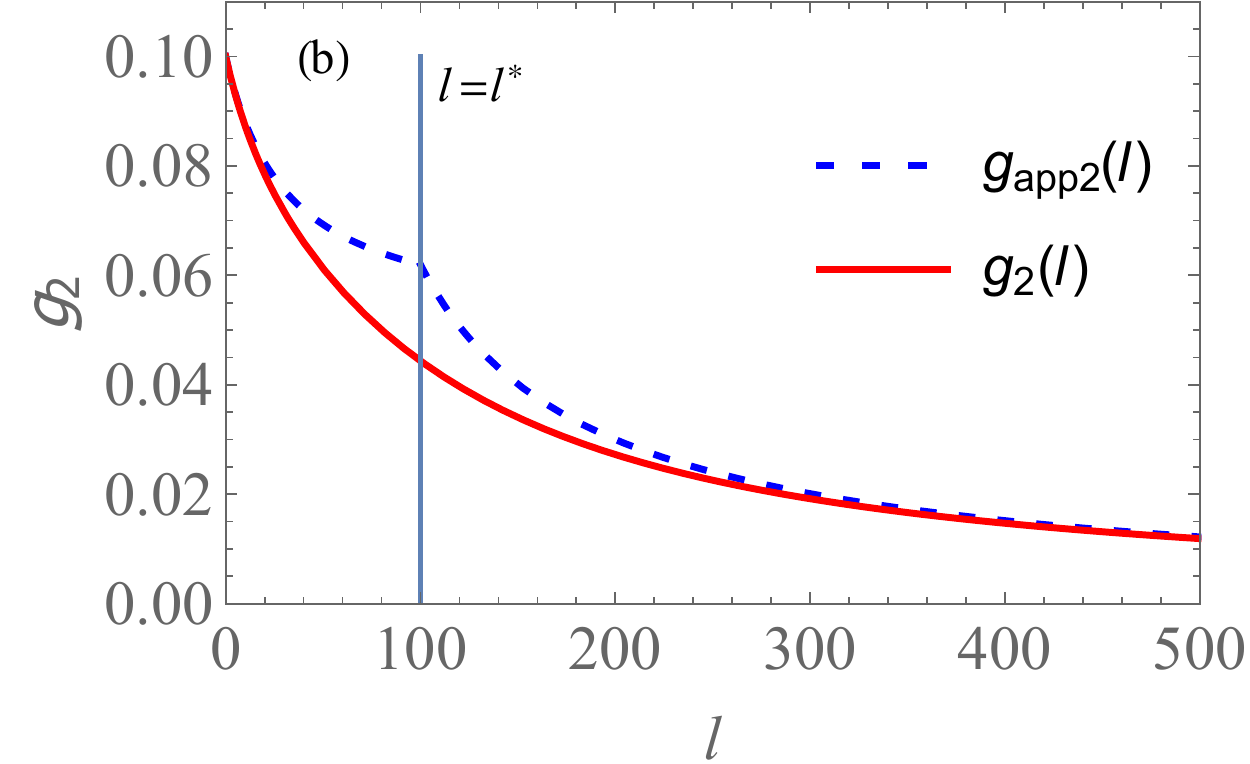,width=0.5\columnwidth}}
  \caption{\label{g1gapp1} (color line) The solid lines in (a) and (b) are plotted by solving numerically Eqs. (\ref{floweqng1}) and (\ref{floweqng2}) in $1+\epsilon$ dimensions with $\epsilon=0.01$, $g_{10}=0.1$ and $g_{20}=0.1$. The dashed line in (a) and (b) are plotted using the approximation model Eqs. (\ref{g1app}) and (\ref{g2app}) respectively. The approximation model captures the crossover behavior for $g_1$ and $g_2$ with a crossover scale $l^*=1/\epsilon$.
  }
  \end{figure}

where $g_{10}$ and $g_{20}$ are the initial values of $g_{1}$ and $g_{2}$ respectively. For $l>>l^*$, we can neglect the exponential terms in Eqs. (\ref{floweqng1}) and (\ref{floweqng2}); we then have the equations~\cite{T1}: $dg_{\pm}/dl=-g_{\pm}^2/(2\pi)$, where $g_{\pm}=g_2\pm g_1$. The solutions are
\beq
g_1(l)=\frac{g_{1c}}{(1+g_{+c}(l-l_c)/(2\pi))(1+g_{-c}(l-l_c)/(2\pi))},
\eeq
\beq
g_2(l)=\frac{g_{2c}+g_{+c}g_{-c}(l-l_c)/(2\pi)}{(1+g_{+c}(l-l_c)/(2\pi))(1+g_{-c}(l-l_c)/(2\pi))},
\eeq
where $g_{1c}$ and $g_{2c}$ are the values of $g_1$ and $g_2$ respectively when entering the $l>>l^*$ region, and $g_{\pm c}=g_{2c}\pm g_{1c}$. We can therefore write down the following \textit{approximate} solutions setting $l_c=l^*=1/\epsilon$, and choosing $g_{1c}=g_1(l_c)$,  $g_{2c}=g_2(l_c)$ using the small $l$ solutions:
\beq
g_{1app}(l)=\theta(l^*-l)\frac{g_{10}}{1+g_{10}l/\pi}+\theta(l-l^*)
\frac{g_{1c}}{(1+g_{+c}(l-l^*)/(2\pi))(1+g_{-c}(l-l^*)/(2\pi))},\label{g1app}
\eeq
\beq
g_{2app}(l)=\theta(l^*-l)(g_{20}-\frac{g_{10}}{2}+\frac{g_{10}}{2(1+g_{10}l/\pi)})+\theta(l-l^*)
\frac{g_{2c}+g_{+c}g_{-c}(l-l^*)/(2\pi)}{(1+g_{+c}(l-l^*)/(2\pi))(1+g_{-c}(l-l^*)/(2\pi))},\label{g2app}
\eeq
where $g_{1c}=g_{10}/(1+g_{10}/(\epsilon\pi))$ and $g_{2c}=g_{20}-g_{10}/2+g_{10}/[2(1+g_{10}/(\epsilon\pi))]$. \figdisp{g1gapp1} (a) and (b)  show a comparison of these approximate solutions for $g_1$ and $g_2$ with the exact (numerical) solutions of Eq.s (\ref{floweqng1}) and (\ref{floweqng2}).

We can see from these figures that $g_{1app}$ and $g_{2app}$ are  fairly good approximations for the exact $g_1$ and $g_2$ in the two asymptotic regions $l<<l^*$ and $l>>l^*$. In 1-d, if $g_{10}=0$, then $g_2(l)=g_{20}$ - this is referred to as the 1-d fixed point model; even when $g_{10} \ne 0$, $g_1(l)\to 0$ and $g_2(l)\to g_{20}-g_{10}/2$ as $l \to \infty$. However these results no longer hold in the case of $1+\epsilon$ dimensions - for  $g_{10}=0$ as well as for almost all other initial conditions, $g_2(l)$ goes to zero asymptotically as $1/(l-l_c)$ for $l>>l^*$.


\section{Calculation of  $Z$, the quasiparticle weight}\label{zz}
\begin{figure}
  \centerline{\epsfig{file=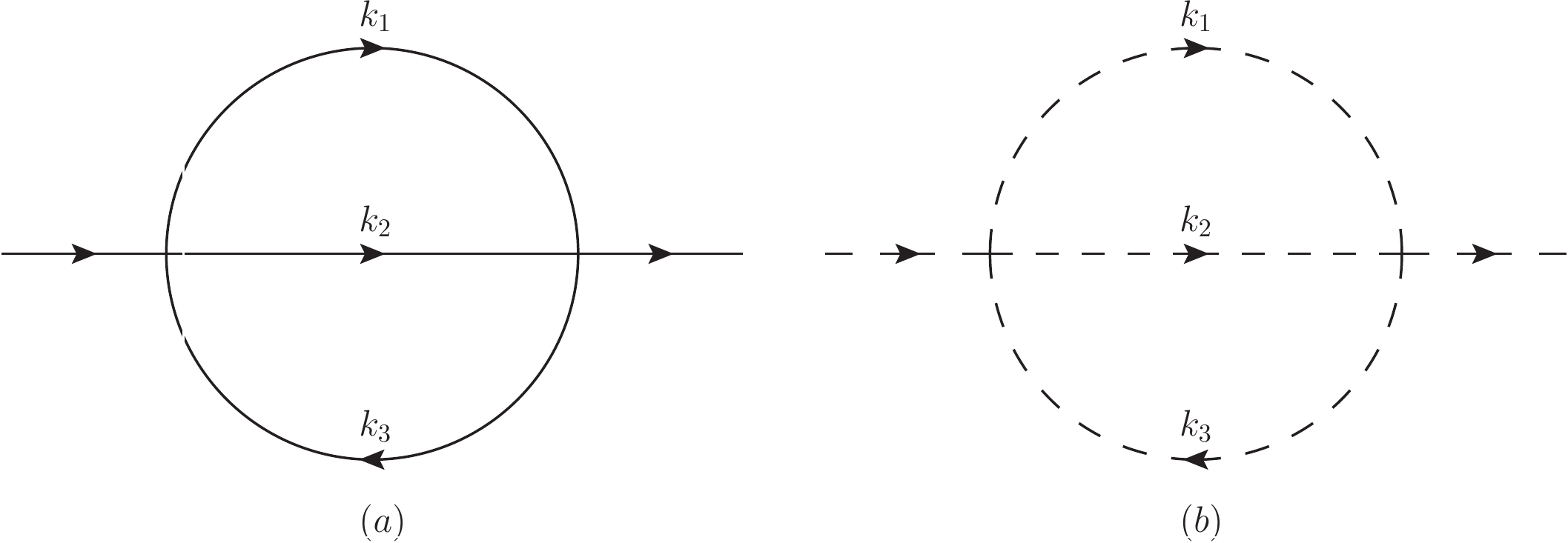,width=0.6\columnwidth}}
  \caption{\label{g422} Sunrise diagrams with all momenta in the left branch (a) or right branch (b).}
\end{figure}

To calculate $Z$, it is convenient to rewrite the interaction terms in  action in \disp{1da} as
\barray
V_{int}[\{\phi\}]&=& g_1 \phi^{*}_{s,L}(1)\phi^{*}_{\bar{s},R}(2) \phi_{\bar{s},L}(3) \phi_{s,R}(4)\Delta
+g_2\phi^{*}_{s,L}(1)\phi^{*}_{\bar{s},R}(2)\phi_{\bar{s},R}(3) \phi_{s,L}(4) \Delta \nn \\
&&+(g_1-g_2) \phi^{*}_{s,L}(1)\phi^{*}_{s,R}(2)\phi_{s,L}(3) \phi_{s,R}(4) \Delta
+\sum_{\alpha=L,R}\frac{g_4}{2}\phi^{*}_{s,\alpha}(1) \phi^{*}_{s',\alpha}(2)\phi_{s',\alpha}(3) \phi_{s,\alpha}(4)\Delta
\earray
where $\bar{s}$ is the opposite spin of $s$. Since we are only interested in $1+\epsilon$ dimensions with $\epsilon\ll 1$, we use the same prescription as in Section \ref{EPMS} in our EPMS calculation of $Z$. Therefore the integrals including the $\delta$ functions look the same as in the 1-d case, except for the additional $|k|^\epsilon$ factor in the Peierls channels. In the second order sunrise diagram, it is easy to see that each of the two body interaction couplings above only couples to itself.  The $g_1^2$, $g_2^2$ and $(g_1-g_2)^2$ terms in the  contributions to the self energy are all given by the same diagram, as shown in \figdisp{sunrise}; so the net contribution is proportional to $g^2 \equiv g_1^2+g_2^2+(g_1-g_2)^2$. Furthermore, there are  contributions to the self energy proportional to $g_4^2$ coming from the diagrams in \figdisp{g422}.

The  calculation of the self energy using the EPMS prescription extended to $1+\epsilon$ dimensions using the sunrise diagrams like the ones shown in \figdisp{sunrise} and \figdisp{g422}, labelled such that  $k_1$ and $k_3$ are from the same branch (left branch for example), will hence involve the integrals
\beq
\begin{split}
&\int_{-\varLambda_0}^{\varLambda_0}\frac{dk_{1}}{2\pi}\int_{-\varLambda_0}^{\varLambda_0}\frac{dk_{2}}{2\pi} \int_{-\varLambda_0}^{\varLambda_0} \frac{|k_3|^\epsilon dk_{3}}{2\pi}\int_{-\infty}^{+\infty} \frac{d\omega_{1}}{2\pi}\int_{-\infty}^{+\infty}\frac{d\omega_{2}}{2\pi}\int_{-\infty}^{+\infty} \frac{d\omega_{3}}{2\pi}\frac{1}{i\omega_1+k_1}\frac{1}{i\omega_2\pm k_2}\frac{1}{i\omega_3+k_3}
\\&\delta(k_1+k_2-k_3-k)\delta(\omega_1+\omega_2-\omega-\omega_3)
\\&=\int_{-\varLambda_0}^{\varLambda_0}\frac{dk_{2}}{2\pi}\int_{-\varLambda_0}^{\varLambda_0} \frac{|k_3|^\epsilon dk_{3}}{2\pi}\int_{-\infty}^{+\infty}\frac{d\omega_{2}}{2\pi}\int_{-\infty}^{+\infty} \frac{d\omega_{3}}{2\pi}\frac{1}{i(\omega_3+\omega-\omega_2)+(k_3+k-k_2)}\frac{1}{i\omega_2\pm k_2}\frac{1}{i\omega_3+k_3}
\\&=\int_{-\varLambda_0}^{\varLambda_0}\frac{dk_{2}}{2\pi}\int_{-\varLambda_0}^{\varLambda_0} \frac{|k_1|^\epsilon dk_{1}}{2\pi}\int_{-\infty}^{+\infty}\frac{d\omega_{2}}{2\pi}\int_{-\infty}^{+\infty} \frac{d\omega_{1}}{2\pi}\frac{1}{i(\omega_1+\omega_2-\omega)+(k_1+k_2-k)}\frac{1}{i\omega_2\pm k_2}\frac{1}{i\omega_1+k_1}, \label{equiv}
\end{split}
\eeq
where in the last step we have set $k_1=-k_3$ and $\omega_1=-\omega_3$. The last result is exactly what we would have obtained by including the $\epsilon$ dependent factor into the $k_1$ integral rather than into the $k_3$ integral at the outset; i.e.,  in the case with $k_1$ and $k_3$ from the same branch, including $|k_3|^\epsilon$ factor into $k_3$ integral is equivalent to including $|k_1|^\epsilon$ into $k_1$ integral. Henceforth, for convenience, we use $|k_1|^\epsilon$ in the following calculations of $Z$ and $\Im m \, {\Sigma}$  in $1+\epsilon$ dimensions comparison with the formalism in the  1-d case. Also, since we are interested in the $\omega$ dependent part of the self energy, we set the external $k=0$ (at the Fermi surface) without loss of generality.

First we calculate the $g_4^2$ diagram in \figdisp{g422}. For the $(n+1)^{th}$ step of EPMS, we need to calculate (compare \disp{III})
\beq
\begin{split}
I_{g4}(k=0,i\omega;\varLambda_n)=&[g_4(\varLambda_n)]^2\int_{-\varLambda_0}^{\varLambda_0}\frac{|k_1|^\epsilon dk_{1n}}{2\pi}\int_{-\varLambda_0}^{\varLambda_0}\frac{dk_{2n}}{2\pi}\int_{-\infty}^{+\infty} \frac{d\omega_{1n}}{2\pi}\int_{-\infty}^{+\infty}\frac{d\omega_{2n}}{2\pi}
\\&\frac{1}{i\omega_{1n}-k_{1n}}\frac{1}{i\omega_{2n}-k_{2n}}\frac{1}{i(\omega_{1n}+\omega_{2n}-\omega_n) -(k_{1n}+k_{2n})},\label{g42}
\end{split}
\eeq
where, as before, $\varLambda_n = \Lambda_0 /s^n =\Lambda_0 / \textrm{e}^{n \, dl}$. The integral will vanish after integrating over $\omega_{1n}$ and $\omega_{2n}$ unless the integrand has poles in different half planes; that is, either $(k_{1n}>0$, $k_{2n}>0$, $k_{1n}+k_{2n}<0)$ or $(k_{1n}<0$, $k_{2n}<0$, $k_{1n}+k_{2n}>0$). Either set of conditions  is impossible to satisfy, so the integral vanishes; hence the  $g_4^2$ term does not contribute to $Z$.

Next, consider the contribution proportional to $g^2$. The relevant integral is
\barray
I(k=0,i\omega;\varLambda_n)&=&[g(\varLambda_n)]^2s^n\int_{-\varLambda_n}^{\varLambda_n}\frac{|k_1|^\epsilon dk_1}{2\pi}\int_{-\varLambda_n}^{\varLambda_n}\frac{dk_2}{2\pi}\int_{-\infty}^{+\infty}\frac{d\omega_1}{2\pi} \int_{-\infty}^{+\infty}\frac{d\omega_2}{2\pi} \nn \\
&&\times\frac{1}{i\omega_1+k_1}\frac{1}{i\omega_2-k_2}\frac{1}{i(\omega_1+\omega_2-\omega)+(k_1+k_2)} \label{Ig2}
\earray
Evaluating the frequency integrals by contour integration, we see that there are two regions of $k_1-k_2$ space which can lead to non-vanishing contributions: either $(k_1>0$, $k_2<0$, $k_1+k_2<0)$ or $(k_1<0$, $k_2>0$, $k_1+k_2>0)$. In the former case, after simplifying we get the condition $(-\varLambda_n<k_2<0$, $0<k_1<-k_2)$, leading to the contribution
\beq
\begin{split}
I_1(k=0, i\omega; \varLambda_n)&=[g(\varLambda_n)]^2s^n\int^{\varLambda_n}_0\frac{dk_2}{2\pi}\int_0^{k_2}\frac{|k_1|^\epsilon dk_1}{2\pi}\frac{-1}{i\omega+2k_2}.\label{I1}
\end{split}
\eeq
Likewise, in the latter case, we have the contribution
\beq
I_2(k=0, i\omega; \varLambda_n)=[g(\varLambda_n)]^2s^n\int^{\varLambda_n}_0\frac{dk_2}{2\pi}\int_0^{k_2} \frac{|k_1|^\epsilon dk_1}{2\pi}\frac{-1}{i\omega-2k_2}.\label{I2}
\eeq
It is convenient at this stage to change into real frequencies by the analytic continuation ($i\omega \rightarrow \omega+i\delta \equiv \omega^+$) so that we are looking at the renormalization of the retarded Green function. Thus, from Eqs. ({\ref{I1}}) and ({\ref{I2}}), we get 
\beq
\begin{split}
I(k=0, \omega^{+}; \varLambda_n)&=I_1(k=0, \omega^+; \varLambda_n)+I_2(k=0, \omega^+; \varLambda_n)
\\&=-[g(\varLambda_n)]^2s^n\int_0^{\varLambda_n}\frac{k_2^\epsilon dk_2}{4\pi^2(1+\epsilon)} (\frac{-\omega/2}{\omega^++2k_2}+\frac{\omega/2}{\omega^+-2k_2}). \label{Ik=0}
\end{split}
\eeq
Therefore, as per the prescription described in Appendix \ref{EPMS}, the contribution to the incremental self energy from the EPMS step reducing the cutoff from $\varLambda_n$ to $\varLambda_{n+1}$ is 
\beq
\begin{split}
\Delta I(k=0, \omega^+; \varLambda_n, \varLambda_{n+1})&=I(k=0, \omega^+; \varLambda_n)-\frac{I(k=0, \omega^+; \varLambda_{n+1})[g(\varLambda_{n})]^2}{s[g(\varLambda_{n+1})]^2}
\\&=-[g(\varLambda_n)]^2s^n\int_{\varLambda_{n+1}}^{\varLambda_n}\frac{k_2^\epsilon dk_2}{4\pi^2(1+\epsilon)}(\frac{-\omega/2}{\omega^++2k_2}+\frac{\omega/2}{\omega^+-2k_2})
\\&=[g(\varLambda_n)]^2\frac{\omega_{n}[(\varLambda_n)^\epsilon-(\varLambda_{n+1})^\epsilon]}
{8\pi^2\epsilon(1+\epsilon)}+o(\omega^3)+i\Im m\Delta I(k=0,\omega; \varLambda_n,\varLambda_{n+1}),\label{deltaI}
\end{split}
\eeq
where $\omega_{n}=s^n\omega$ as in Appendix \ref{Shankar-RG}. We note that the self energy contribution is purely real unless $\varLambda_{n+1}<|\omega|/2<\varLambda_n$, and when this condition is satisfied, we have
\beq
\Im m\Delta I(k=0, \omega; \varLambda_n, \varLambda_{n+1})=\frac{[g(\varLambda_n)]^2s^n}{8\pi^2(1+\epsilon)}|
\frac{\omega}{2}|^{1+\epsilon}.\label{ImdeltaI}
\eeq

Noting from its definition that incremental contributions to  $Z$ accumulate  multiplicatively in the same way  as  for the multiplicative renormalization factor $a^{-1}$ in \disp{Sphiless}, and following \disp{arelation}, we can calculate the $Z$ factor  after $n$ steps of EPMS as the product
\beq
Z(\varLambda_n)=\prod_{m=0}^{n-1} \tilde{Z}(\varLambda_m \rightarrow \varLambda_{m+1}).\label{zprod}
\eeq
Here $[\tilde{Z}(\varLambda_m \rightarrow \varLambda_{m+1})]^{-1}-1$ is the lowest order contribution to the coefficient of $\omega$ in the   real part of the self energy arising from the $(m+1)^{th}$ EPMS step reducing the cutoff from $\varLambda_m$ to $\varLambda_{m+1}$.
We note from the above that $\ln(Z)$ is the \textit{sum} of  \textit{additive} incremental contributions from each step of EPMS. Hence by making these steps infinitesimal
as before by the choice $s=\textrm{e}^{dl}$,  we can derive a differential equation for $\ln(Z)$: 
\beq
\begin{split}
\tilde{Z}(\varLambda_m \rightarrow \varLambda_{m+1})=\frac{1}{1+\frac{\partial\re\Delta I(k=0, \omega; \varLambda_m, \varLambda_{m+1})}{\partial \omega_{n}}\Big|_{\omega_{n}\rightarrow 0}}=1-\frac{g^2(\varLambda_m)\textrm{e}^{-\epsilon mdl}}{8\pi^2(1+\epsilon)}dl;
\end{split}
\eeq
whence, keeping $n \, dl=l$ and $\varLambda_n = \textrm{e}^{-n \, dl} = \textrm{e}^{-l}$ fixed while letting $n \to \infty$ and $dl \to 0$
\beq
d\ln Z_l=\ln \tilde{Z}(\varLambda_n \rightarrow \varLambda_{n+1})=\ln (1-\frac{g_l^2\textrm{e}^{-\epsilon n \, dl}}{8\pi^2(1+\epsilon)}dl)=-\frac{g_l^2\textrm{e}^{-\epsilon l}}{8\pi^2(1+\epsilon)}dl,
\eeq
where we have denoted  $Z_l \equiv Z(\varLambda_n)$ and $g_l^2 \equiv [g(\varLambda_n)]^2$. Thus we get the differential equation for $\ln Z$,
\beq
\frac{d\ln Z}{dl}=-\frac{g_l^2\textrm{e}^{-\epsilon l}}{8\pi^2(1+\epsilon)},\label{floweqnlnz}
\eeq
By definition\cite{Kopietz}, the flowing \textit{anomalous dimension} is therefore
\beq
\eta(l)\equiv -\frac{d\ln Z}{dl}=\frac{g_l^2\textrm{e}^{-\epsilon l}}{8\pi^2(1+\epsilon)}.
\eeq
The emergent crossover  scale $l^*$ is evident here. From $g_l^2 = 2(g_1^2+g_2^2-g_1g_2) = 2 [(g_2-g_1/2)^2+3g_1^2/4]$ and using Eqs. (\ref{g1app}) and (\ref{g2app}) in Appendix \ref{g1g2}, it is easy to see that for $1 \ll l \ll l^*$, $g_l^2  \approx 2(g_{20}-g_{10}/2)^2$ and for $l\gg l^*$, $g_l^2\approx 8\pi^2/l^2$. Hence, when $1 \ll l \ll l^*$, the anomalous dimension is essentially the same as in the  1-d case: $\eta(l)\approx [g_{20}-g_{10}/2]^2/(4\pi^2)$. When $l\gg l^*$, $\eta(l)$ or $d\ln Z/dl\to0$, hence the anomalous dimension vanishes and $Z$ converges to a constant in  $1+\epsilon$ dimensions. From \disp{floweqnlnz}, we have
\beq
\ln Z_l=-\int_0^l\frac{[g(l')]^2\textrm{e}^{-\epsilon l'}}{8\pi^2(1+\epsilon)}dl'. \label{Zintegration}
\eeq\begin{figure}
  \centerline{\epsfig{file=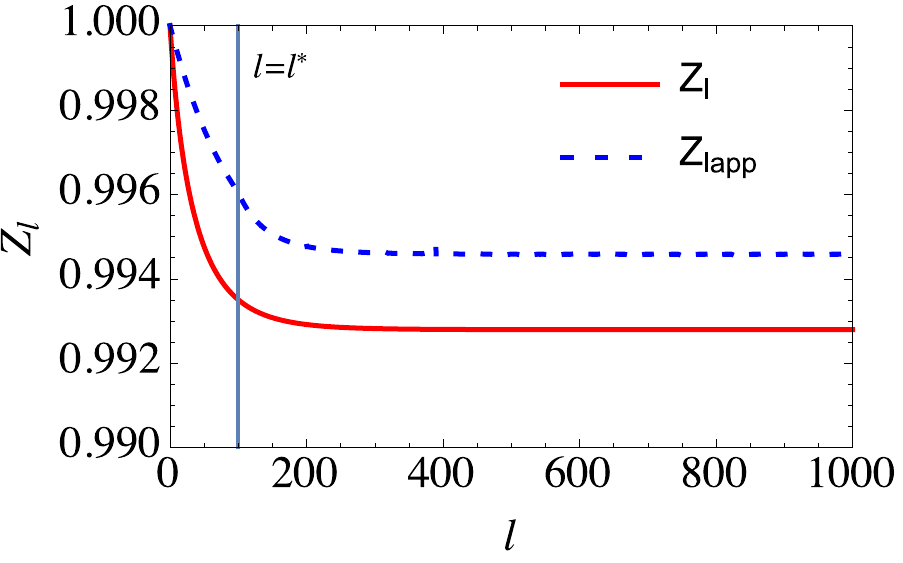,width=0.4\columnwidth}}
  \caption{\label{zzapp} (color line) The red solid line represents exact numerical $Z_l$ obtained first solving Eqs. (\ref{floweqng1}) and (\ref{floweqng2}) with $\epsilon=0.01$, $g_{10}=0.1$ and $g_{20}=0.1$, and then substituting the results into $g_{l'}^2$ in Eq. (\ref{Zintegration}). The dashed blue line represents the approximate analytical model $Z_{lapp}$ in Eq. (\ref{zapp}) which captures the crossover behaviors in region $l\gg l^*$ and $l\ll l^*$ where $l^*=1/\epsilon$ is the crossover scale. When $l\ll l^*$, $Z_l$ decays very fast like the 1-d case, but converges a finite value when $l\gg l^*$ as a feature in 1+$\epsilon$ dimension. 
  }
\end{figure}
From the above results for $g_l^2$, we can write down the following approximate model which permits an analytic calculation of $Z$:
\beq
[g_{lapp}]^2=\theta(l^*-l)2(g_{20}-\frac{g_{10}}{2})^2+\theta(l-l^*)\frac{8\pi^2}{l^2}.\label{g^2}
\eeq
This model captures the asymptotic behaviors of $g_1$ and $g_2$ in region $l\ll l^*$ and $l\gg l^*$ as shown in Eqs. (\ref{g1app}) and (\ref{g2app}).  
Therefore, we get
\beq
\begin{split}
Z_{lapp}&\approx\theta(l^*-l)\exp[-\frac{\eta}{1+\epsilon}\frac{1-\textrm{e}^{-\epsilon l}}{\epsilon}]
\\&+\theta(l-l^*)\exp[-\frac{\eta}{1+\epsilon}\frac{1-\textrm{e}^{-1}}{\epsilon} -\frac{\epsilon }{1+\epsilon}(\textrm{e}^{-1}-\textrm{e}^{-\epsilon l}+\textrm{Ei}(-1)-\textrm{Ei}(-\epsilon l))],\label{zapp}
\end{split}
\eeq
 where $\eta=(g_{20}-g_{10}/2)^2/(4\pi^2)$ is the asymptotic value of $d \ln Z/dl$ in 1-d when $g_{10}\neq 0$ or the anomalous dimension when $g_{10}=0$. Here $\textrm{Ei}(x)$ is the exponential function  $\textrm{Ei}(x)=-\int^{\infty}_{x}\frac{\textrm{e}^{-t}}{t}dt$. 

\figdisp{zzapp} shows a comparison of the exact numerical evaluation of $Z_l$ with this approximate analytical result. The exact numerical $Z_l$ is obtained by first solving Eqs. (\ref{floweqng1}) and (\ref{floweqng2}), substituting the results into $g_{l'}^2$ in Eq. (\ref{Zintegration}) and then doing the integration numerically.  The model captures the crossover of $Z$ between the two regimes but is not very accurate in capturing the asymptotic value of $Z$ when $l\rightarrow \infty$. 
 

In order to find the dependence of $Z_{l\rightarrow \infty}$ on $\epsilon$ and coupling constants, we first look at the special case with $g_{10}=0$. When $g_{10}=0$, $g_1(l)$ remains 0 according to Eq. (\ref{floweqng1}). And Eq. (\ref{floweqng2}) becomes
\beq
\frac{dg_2}{dl}=-\frac{g_2^2(1-\textrm{e}^{-\epsilon l})}{2\pi}, 
\eeq
which can be solved analytically as
\beq
g_2(l)=\frac{g_{20}}{1+g_{20}(l-1/\epsilon+\textrm{e}^{-\epsilon l}/\epsilon)/(2\pi)}.
\eeq
Substituting $g_1(l)$ and $g_2(l)$ into Eq. (\ref{Zintegration}), we get
\beq
\ln Z_{\infty}=-\int_0^{\infty}\frac{\textrm{e}^{-\epsilon l'}}{(1+\epsilon)[1/\eta^{0.5}+l-1/\epsilon+\textrm{e}^{-\epsilon l'}/\epsilon]^2}dl',
\eeq
where $\eta=g_{20}^2/(4\pi^2)$ is the anomalous dimension in 1-d fixed point. Since we look at the small $\epsilon$ behavior, we keep only the leading order of $\epsilon$. First replace $1+\epsilon$ by 1 in the denominator. The integral can be separated into two parts $l'<1/\epsilon$ and $l'>1/\epsilon$. For the first part, $\textrm{e}^{-\epsilon l'}\approx 1$ in the numerator and $l-1/\epsilon+\textrm{e}^{-\epsilon l'}/\epsilon\approx \epsilon l'^2/2$ in the denominator. So we have
\beq
\int_0^{1/\epsilon}\frac{1}{[1/\eta^{0.5}+\epsilon l'^2/2]^2}dl'=(\frac{2}{\epsilon})^{0.5}\eta^{3/4}\int_0^{\eta^{0.25}(2/\epsilon)^{0.5}}\frac{1}{[1+l'^2]^2}dl' \approx 1.11 \frac{\eta^{3/4}}{\sqrt{\epsilon}}.\label{firstpart}
\eeq
In the last step, we approximate the upper limit by infinity because we are interested in small $\epsilon$ and the integral converges fast, and then use $\int_0^{\infty}1/(1+l'^2)^2dl'\approx 0.785$. The second part of the integral is
\beq
\int_{1/\epsilon}^{\infty}\frac{\textrm{e}^{-\epsilon l'}}{[1/\eta^{0.5}+l'-1/\epsilon+\textrm{e}^{-\epsilon l'}/\epsilon]^2}dl'= \epsilon\int_{1}^{\infty}\frac{\textrm{e}^{-l'}}{[\epsilon/\eta^{0.5}+l'-1+\textrm{e}^{-l'}]^2}dl'=o(\epsilon).
\eeq

\begin{figure}
  \centerline{\epsfig{file=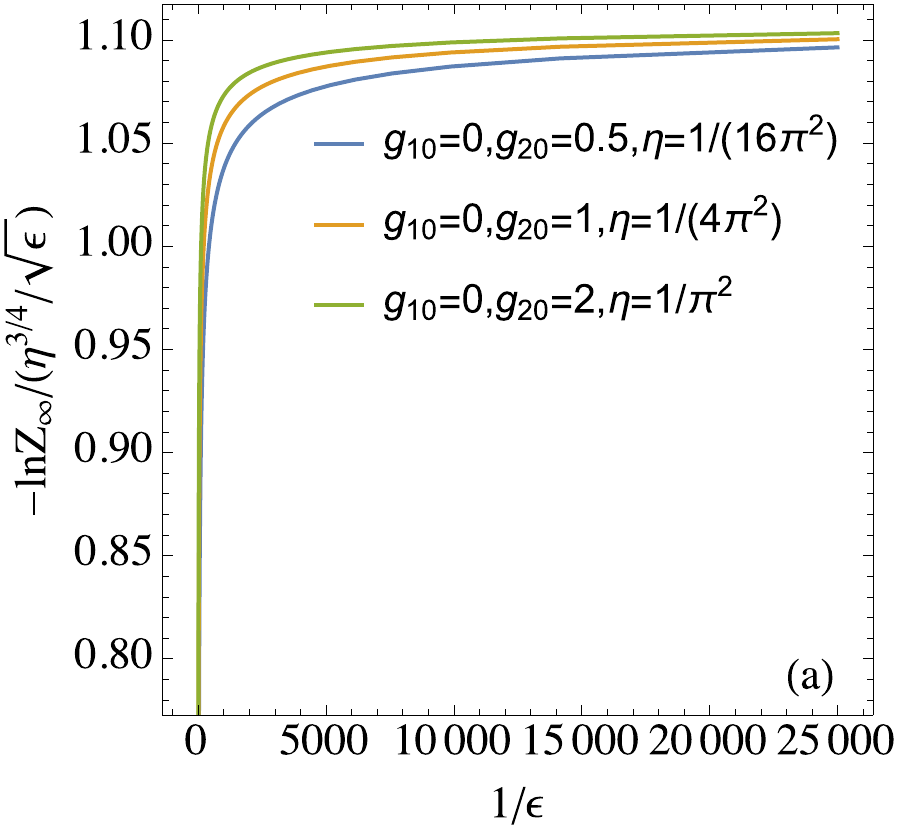,width=0.4\columnwidth}\epsfig{file=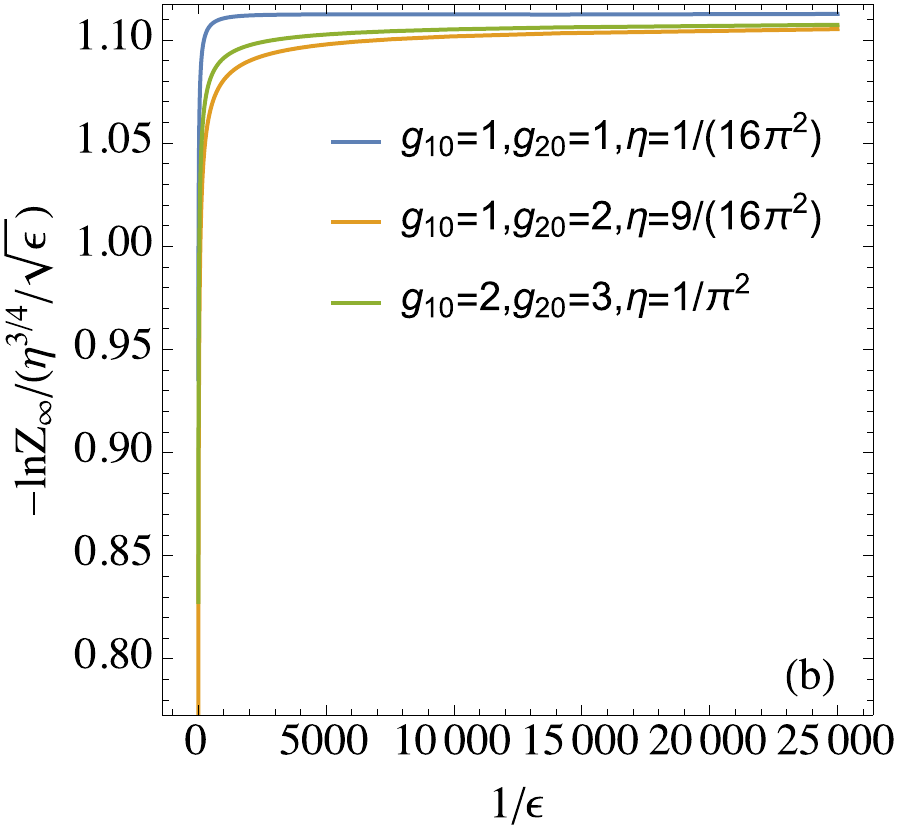,width=0.4\columnwidth}}
  \caption{\label{zomega2} (color line) (a) $-\ln Z_\infty/(\eta^{3/4}/\sqrt{\epsilon})$ versus $1/\epsilon$ is plotted for $g_{10}=0$ and $g_{20}=0.5,\,1,\,2$ corresponding to $\eta=1/(16\pi^2)),\,1/(4\pi^2),\,1/\pi^2$ respectively to show that $Z_\infty$ vanishes as $\exp[-d_0 \,\eta^{3/4}/\sqrt{\epsilon}]$ with $d_0\approx 1.09$ when $\epsilon\rightarrow 0$. (b) $-\ln Z_\infty/(\eta^{3/4}/\sqrt{\epsilon})$ versus $1/\epsilon$ is plotted for several combination of $g_{10}\neq 0$ and $g_{20}$. Three curves also converge to about 1.09. It shows that in the general case, $Z_\infty$ also vanishes as $\exp[-d_0 \,\eta^{3/4}/\sqrt{\epsilon}]$ with $d\approx 1.09$ when $\epsilon\rightarrow 0$.
  }
\end{figure}

Therefore we can keep only the first part of the integral when $\epsilon$ is very small and 
\beq
Z_{\infty}=\exp\{- d_0\frac{{\eta}^{3/4}}{\sqrt{\epsilon}}\},  \label{zzero}
\eeq
where $d_0$ is about 1.11 from \dispeq{firstpart}. The numerical result in \figdisp{zomega2} (a) attests this form with a slightly smaller $d$ as 1.09 because we overestimate the first part of the integral a bit in the analytical calculation. 

Then It is natural to ask whether \dispeq{zzero} is still valid when $g_{10}\neq 0$. Since Eqs. (\ref{floweqng1}) and (\ref{floweqng2}) cannot be solved analytically on a general initial condition, we have to rely on numerical calculation. \figdisp{zomega2} (b) does show that \dispeq{zzero} also works for $g_{10}\neq 0$ with $d_0$ around $1.09$ and $\eta=(g_{20}-g_{10}/2)^2/(4\pi^2)$.

\section{Calculation of the leading $\omega$ dependence of  $Z(\omega)$ and $ \Sigma(\omega)$\label{sigmaomega}}
\begin{figure}
  \centerline{\epsfig{file=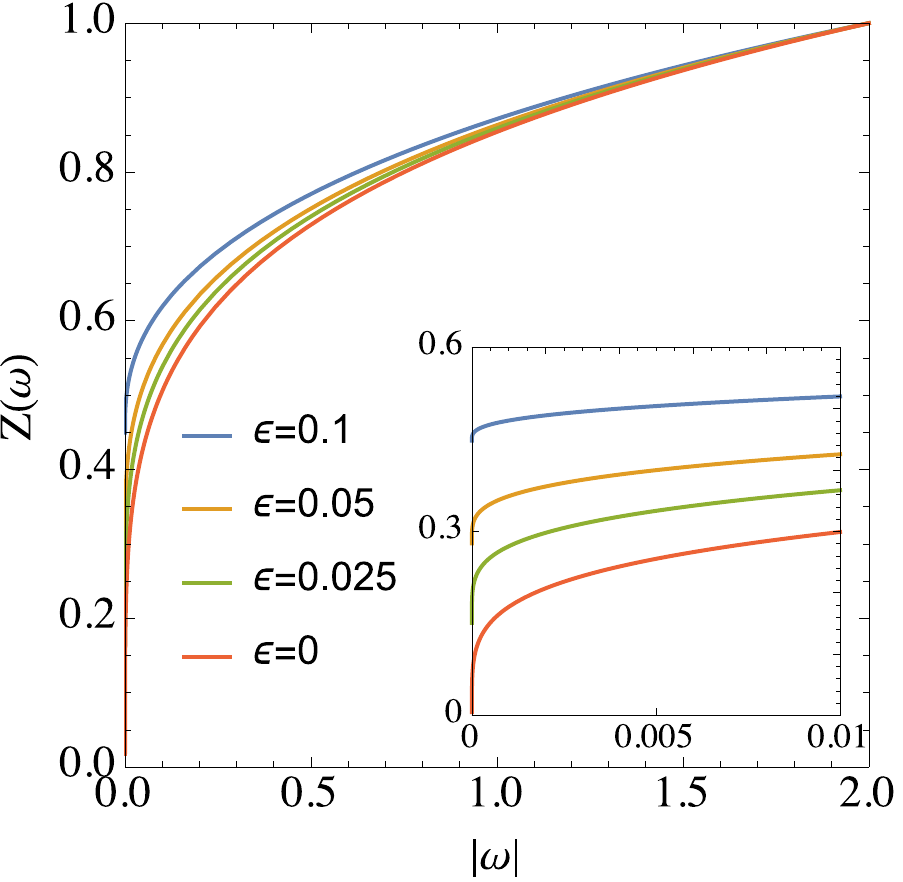,width=0.4\columnwidth}}
  \caption{\label{zomega} (color line) The quasi-particle weight $Z(\omega)$ is plotted for different value of $\epsilon$ with $g_{10}=0$ and $g_{20}=3$ corresponding to the 1-d fixed point $g_{10}=0$. The inset shows $Z(0)$ remains finite for any finite $\epsilon$,  and vanishes when $\epsilon\to 0$.  }
\end{figure}

In this section, we calculate the full, frequency dependent self energy $  \Sigma(\omega)$. In this and next section, we use $\omega$ corresponding to the analytic continuation to real frequencies, $i\omega\to\omega+i\delta \equiv \omega^+$.  We choose $n$ such that $\varLambda_{n+1}<|\omega/2|<\varLambda_n$, for, as seen in the previous section, if this condition is satisfied the leading contribution to the self energy is purely real up to the $n^{th}$ step of EPMS, and its effects on the Greens function are basically captured by $Z$. To recapitulate, after $n$ steps of EPMS, the lowest order (i.e., free) Green function of the \textit{rescaled} fields, $G_n(\varLambda_0)$ (also see Appendix \ref{Shankar-RG}) is
\beq
G_n(\omega_n^+;\varLambda_0)=\frac{1}{\omega_n^+},
\eeq
where we have set the  external $k=0$ (and also suppressed it as an argument of $G$), and kept only the leading term in $\omega$. As before, $\omega_n \equiv s^n\omega $. Hence the low energy \textit{effective} or \textit{renormalized} Green function  of the original fields, but   with the reduced cutoff, is, to leading order,
\beq
G(\omega^+; \varLambda_n)=s^nZ_{n \, dl}G_n(\varLambda_0)=\frac{Z_l}{\omega^+},
\eeq
which is a restatement of \disp{gfrelation} in the Appendix \ref{Shankar-RG} using $Z$ instead of $a$. We keep going back to the original leading order Green function with a reduced cutoff because one aim of RG is to be able to calculate the correlation function of slow modes in a low energy effective theory~\cite{Shankar}.  However, again as shown in the previous section, because of the chosen relation between $\omega$ and $n$, during the next step of EPMS, according to Eqs. (\ref{deltaI}) and (\ref{ImdeltaI}), we get a non-trivial self energy, with an imaginary part, leading to the Green function:
\beq
G_{n+1}(\omega_{n+1}; \varLambda_0)=\frac{1}{s^{n+1}\omega+i\tilde{Z}(\varLambda_n\rightarrow \varLambda_{n+1})\frac{g_l^2s^{n+1}}{8\pi^2(1+\epsilon)}|\frac{\omega}{2}|^{1+\epsilon}},
\eeq
where, as before, we use $-\ln |\omega/2| \approx \ln\varLambda_n=n \, dl$ as the argument of the running coupling constant, and will eventually take the limit $dl \rightarrow 0$. The low energy effective Green function is therefore
\beq
G(\omega; \varLambda_{n+1})=Z_{(n+1)dl}s^{n+1}G_{n+1}(\omega_{n+1};\varLambda_0) \approx \frac{1}{Z_{(n+1)dl}^{-1}\omega+i\frac{g_l^2}{8\pi^2(1+\epsilon)}Z_{n \, dl}^{-1}|\frac{\omega}{2}|^{1+\epsilon}},\label{legf}
\eeq
Now we take the limit $dl \rightarrow 0$ and $n\rightarrow \infty$, fixing $n \, dl=l=-\ln|\omega/2|$. And then we can replace the dependence on $l$ by $\omega$. The Green function is
\beq
G(k_F,\omega)=\frac{1}{[Z(\omega)]^{-1}\omega+i\frac{[g(\omega)]^2}{8\pi^2(1+\epsilon)}[Z(\omega)]^{-1}|\frac{\omega}{2}|^{1+\epsilon}}, \label{fullG}
\eeq
where $[g(\omega)]^2=g_l^2$ with $l=-\ln |\omega/2|$ and  
\beq
Z(\omega)=Z_l=\exp[- \frac{1}{2^\epsilon} \, \int_\omega^2\frac{[g(\omega')]^2  {\omega'}^{\epsilon-1}}{8\pi^2(1+\epsilon)}d\omega'], \label{Z1}
\eeq
where the upper limit "2" is the dimensionless bandwidth. We may rewrite this usefully as
\beq
Z(\omega)=Z(0) \times \exp[ \frac{1}{2^\epsilon} \, \int_0^\omega\frac{[g(\omega')]^2 {\omega'}^{\epsilon-1}}{8\pi^2(1+\epsilon)}d\omega'], \label{Z2}
\eeq
where $Z(0)$ is found from \disp{Z1} by extending the lower integral to 0, and we note its value below. To compare with the fix point ($g_{10}=0$) model in 1-d, we plot $Z(\omega)$ versus $|\omega|$ in  \figdisp{zomega} for several values of $\epsilon,$ with $g_{10}=0$. In the fixed point model in 1-d, there is an anomalous dimension $\eta=g_{20}^2/(4\pi^2)$,  and when $\omega\to 0$, $Z\to 0$. In $1+\epsilon$ dimensions, when $\omega\to 0$, $Z\to \exp(-d_0\,\eta^{3/4}/\sqrt{\epsilon})$, as discussed in Section \ref{zz}. So when $\epsilon<\eta$, $Z$ is considerably smaller than $1$. We can define the system in this regime,  with a very small $Z$, as  a fragile Fermi Liquid.

We can invert \disp{Z2}  to express 
\beq
[g(\omega)]^2= 2^\epsilon (1+ \epsilon) 8 \pi^2 \; \; \frac{Z'(\omega)}{Z(\omega)} \times  \omega^{1-\epsilon}. \label{gintermsofz}
\eeq
This gives us leading low energy behavior
\beq
G^{-1}(k_f,\omega)|_{\omega\to 0}  \, \sim\frac{\omega}{Z(0)} + \frac{i }{2}\,  Z'(\omega) \times \left( \frac{\omega}{Z(0)}\right)^2. \label{gsimple}
\eeq
Notice the similarity with \disp{damping-g} where the Fermi liquid Greens function is noted in other interesting  cases. In particular, if $Z'(\omega)$ were finite at $\omega\to0$, this would be similar to a standard Fermi liquid quasiparticle Greens function including the leading damping term. However we see next that $Z'$ diverges at the lowest energies as
\beq
Z'(\omega)\sim Z(0) \times\frac{1}{2^\epsilon(1+\epsilon)} \, \frac{1}{\omega^{1-\epsilon} \, (\log(\omega/2))^2}.
\eeq
This changes the damping rate from the familiar quadratic in $\omega$ to $\omega^{1+\epsilon}/ (\log(\omega/2))^2$. Combining with \disp{gsimple} we obtain:
\beq
G^{-1}(k_f,\omega)|_{\omega\to 0}  \, \sim\frac{\omega}{Z} + \frac{i }{2^{1+\epsilon}(1+\epsilon)}\, \left( \frac{|\omega|}{Z}\right)^{1+\epsilon}  \times \frac{1}{(\log(|\omega|/2))^2},
 \label{gsimple2}
\eeq
where $Z=Z(0)$ and we set $Z^\epsilon\to 1$. This expression displays the $\omega/Z$ scaling,  ignoring  the weak  logarithmic term for this purpose.

For small $\omega\ll\textrm{e}^{-1/\epsilon}$ or $\ln|2/\omega|\gg1/\epsilon$, we have 
\beq
\begin{split}
Z(\omega\ll\textrm{e}^{-1/\epsilon})
&\approx\exp(-d_0\frac{\eta^{3/4}}{\sqrt{\epsilon}})\exp[\int_0^\omega\frac{\big|\frac{\omega'}{2}\big|^{\epsilon-1}}{2(\ln|\omega'/2|)^2(1+\epsilon)}d\omega']
\\&=\exp(-d_0\frac{\eta^{3/4}}{\sqrt{\epsilon}})\exp[\frac{\big|\frac{\omega'}{2}\big|^{\epsilon}}{(\ln|\omega'/2|)^2(1+\epsilon)\epsilon}+o(\frac{\big|\omega'\big|^{\epsilon}}{(\ln|\omega|)^3})]
\\&\approx \exp(-d_0\frac{\eta^{3/4}}{\sqrt{\epsilon}})[1+\frac{\big|\frac{\omega'}{2}\big|^{\epsilon}}{(\ln|\omega'/2|)^2(1+\epsilon)\epsilon}+o(\frac{\big|\omega'\big|^{\epsilon}}{(\ln|\omega|)^3})]
\end{split}
\eeq
where we use the small $\omega$ or large $l$ asymptotic behavior of $[g(\omega)]^2$ in Eq.(\ref{g^2}) and integrate by part in the second step. 

\begin{figure}
  \centerline{\epsfig{file=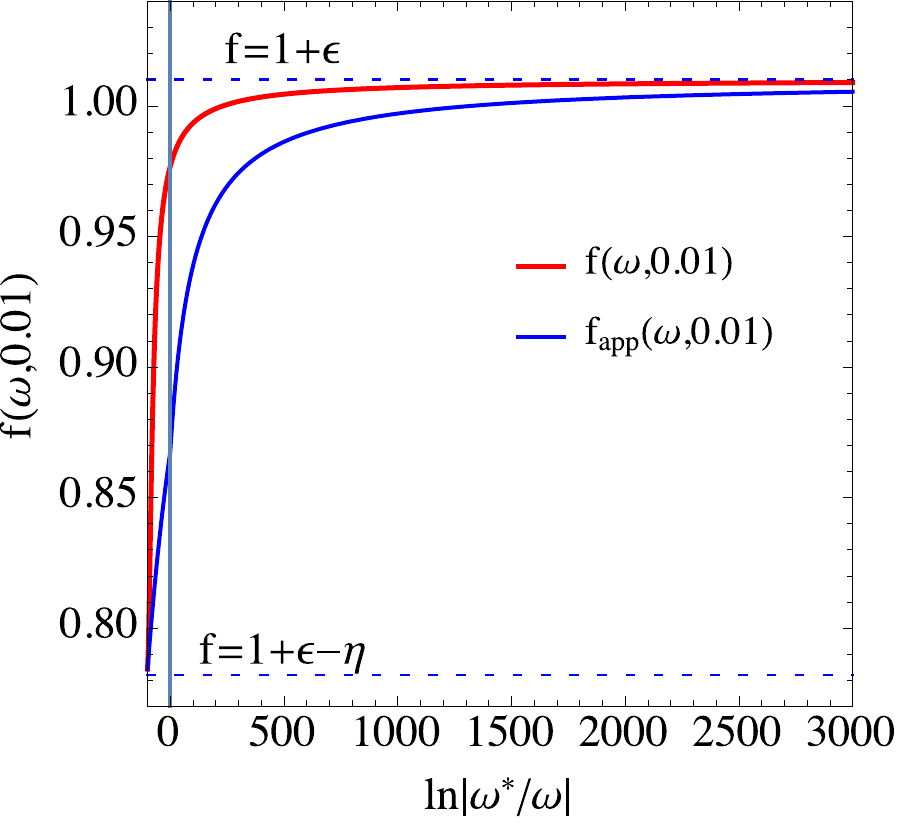,width=0.4\columnwidth}}
  \caption{\label{ffapp} (color line) The red line is plotted by substituting exact numerical $Z(\omega)$ into Eq. (\ref{fff}) with $g_{10}=0$ and $g_{20}=3$ ($\eta=9/(4\pi^2)\approx 0.228$), while the blue line represents the approximation model (\ref{fapp}). This model captures the crossover behavior from Tomonaga-Luttinger liquid region ($\omega\gg\omega^*$) to Fermi liquid region ($\omega\ll\omega^*$) with a crossover scale $\omega^*=2 \textrm{e}^{1/\epsilon}$ and the right asymptotic behaviors in these two regions.
  }
\end{figure}

From \disp{fullG}, we also get the imaginary part of the self-energy at the Fermi surface
\beq
\Im m\Sigma(k_F,\omega)=\Im m \Sigma(k=0,\omega)=-\frac{[g(\omega)]^2}{8\pi^2(1+\epsilon)} [Z(\omega)]^{-1}|\frac{\omega}{2}|^{1+\epsilon}.
\eeq
This can be rewritten as
\beq
\Im m\Sigma(k_F,\omega)=-\frac{[g(\omega)]^2}{8\pi^2(1+\epsilon)}\big|\frac{\omega}{2}\big|^{f(\omega, \epsilon)},\label{imsigma-leading}
\eeq
with 
\beq
f(\omega, \epsilon) \equiv 1+\epsilon -\ln[Z(\omega)]/\ln|\omega/2|.\label{fff}
\eeq 

\begin{figure}
  \centerline{\epsfig{file=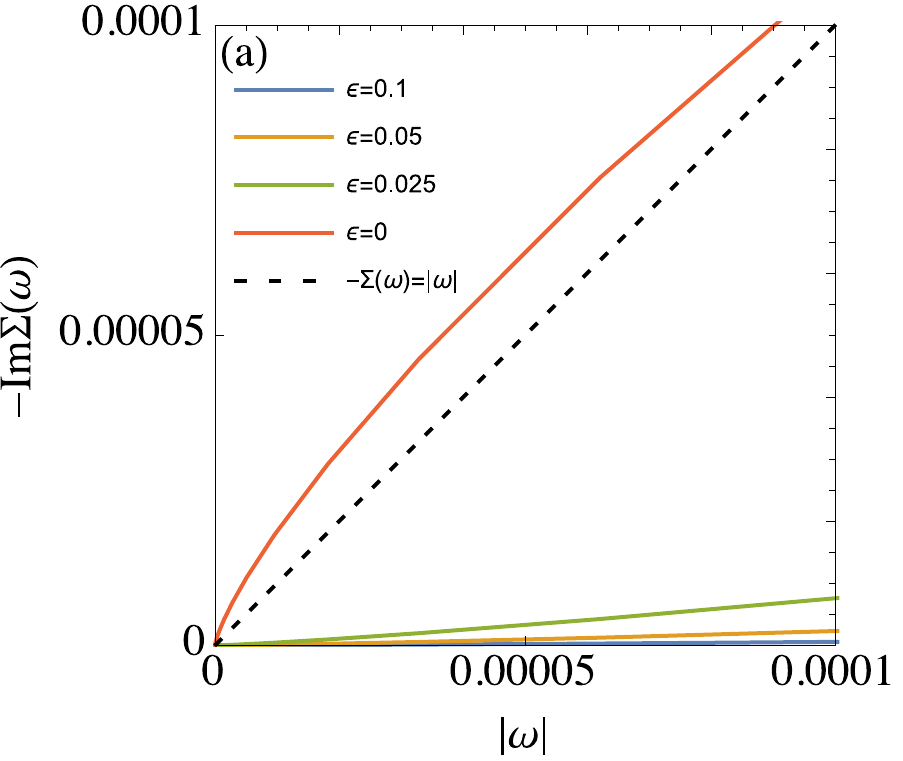,width=0.4\columnwidth}\epsfig{file=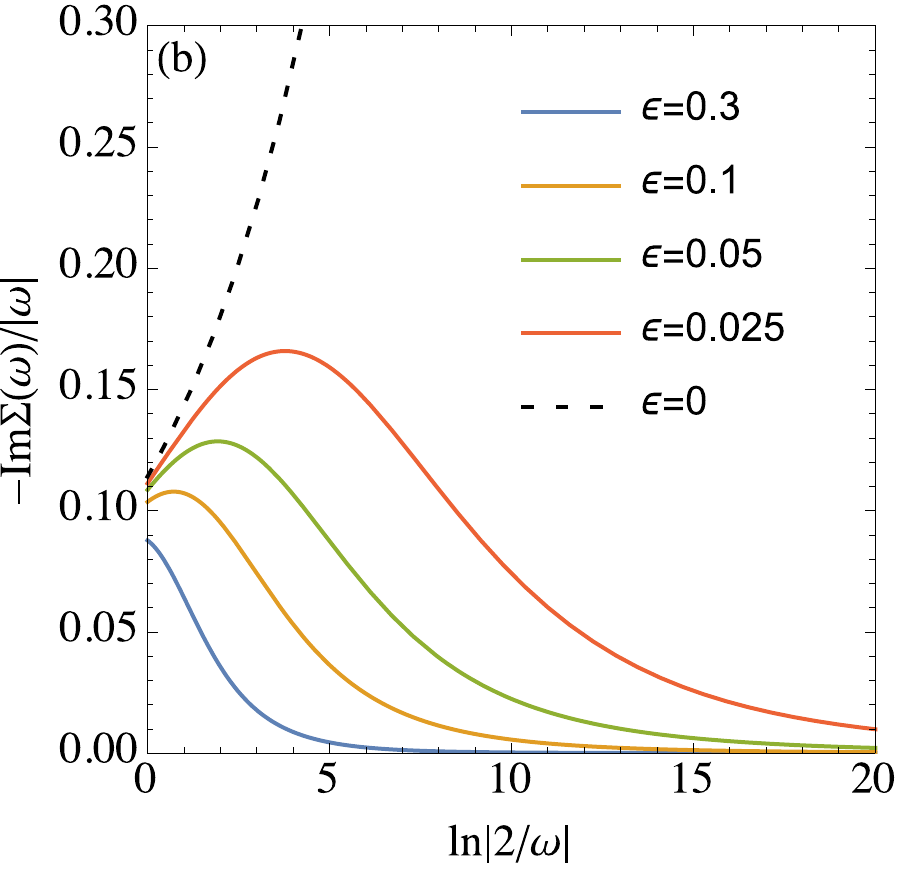,width=0.35\columnwidth}}
  \caption{\label{imsigma}  (color line)  (a) $-\Im m \, \Sigma$ versus $|\omega|$ is plotted for several values of $\epsilon$ with $g_{10}=0$ and $g_{20}=3$. The quasi-particle can be defined only when the decays rate $-\Im m \, \Sigma$ of quasi-particles is much smaller than their energy $\omega$, which is the case when $\epsilon>0$. (b) $-\Im m \, \Sigma$ versus $|\omega|$ is plotted for several values of $\epsilon$ with $g_{10}=0$ and $g_{20}=3~(\eta\approx 0.228)$. When $\epsilon<\eta$, there is an initial increase for large $\omega$, a Non-Fermi liquid behavior like 1-d Tomonaga-Luttinger case, which is a feature of fragile Fermi liquid. For $\epsilon>\eta$, the plot decreases as Fermi liquid behavior from the beginning.
  }
\end{figure}

Using our approximate model $Z_{app}(\omega)$ (which is $Z_{lapp}$ in Eq. (\ref{zapp}) evaluated at $l=-\ln|\omega/2|$) to replace $Z(\omega)$, we get an approximation model 
\beq
\begin{split}
f_{app}(\omega, \epsilon)&=1+\epsilon-\theta(\omega-\omega^*)\frac{\eta}{1+\epsilon}\frac{|\omega/2|^{\epsilon}-1} {\epsilon\ln|\omega/2|}\\&-\theta(\omega^*-\omega)[\frac{\eta}{1+\epsilon}\frac{\textrm{e}^{-1}-1}{\epsilon \ln|\omega/2|}+\frac{\epsilon }{(1+\epsilon)\ln|\omega/2|}(|\omega/2|^{\epsilon}-\textrm{e}^{-1}+\textrm{Ei}(\epsilon\ln|\omega/2|)-\textrm{Ei}(-1))].\label{fapp}
\end{split}
\eeq
where we have used $\textrm{e}^{-l}=|\omega/2|$, $\omega^*=2\textrm{e}^{-1/\epsilon}$. From the formula above, we see $f \to 1+\epsilon-\eta$, corresponding to  1-d like behavior when $\omega\gg\omega^*$ and $f\to1+\epsilon$ when $\omega\ll\omega^*$ as expected for a Fermi liquid in $1+\epsilon$ dimensions. We plot $f(\omega,0.01)$ to show the crossover behaviors in \figdisp{ffapp} and the approximate model $f_{app}(\omega, 0.01)$ captures the right asymptotical behaviors and the crossover. To compare directly with the fixed point model in 1-d, we plot $\Im m\Sigma$ for small frequencies for several values of $\epsilon$ in \figdisp{imsigma} (a). From the asymptotic behaviors of $f$, we see that there is Non-Fermi liquid behavior at relatively high frequencies when $\epsilon<\eta$, which could be the signature of a fragile Fermi liquid. To show this signature, we plot $\Im m\Sigma/\omega$ for  several values of  $\epsilon$ in \figdisp{imsigma} (b). For $\epsilon<\eta$, there is an initial part in the curve which is Non-Fermi liquid behavior, like in 1-d.
\begin{figure}
  \centerline{\epsfig{file=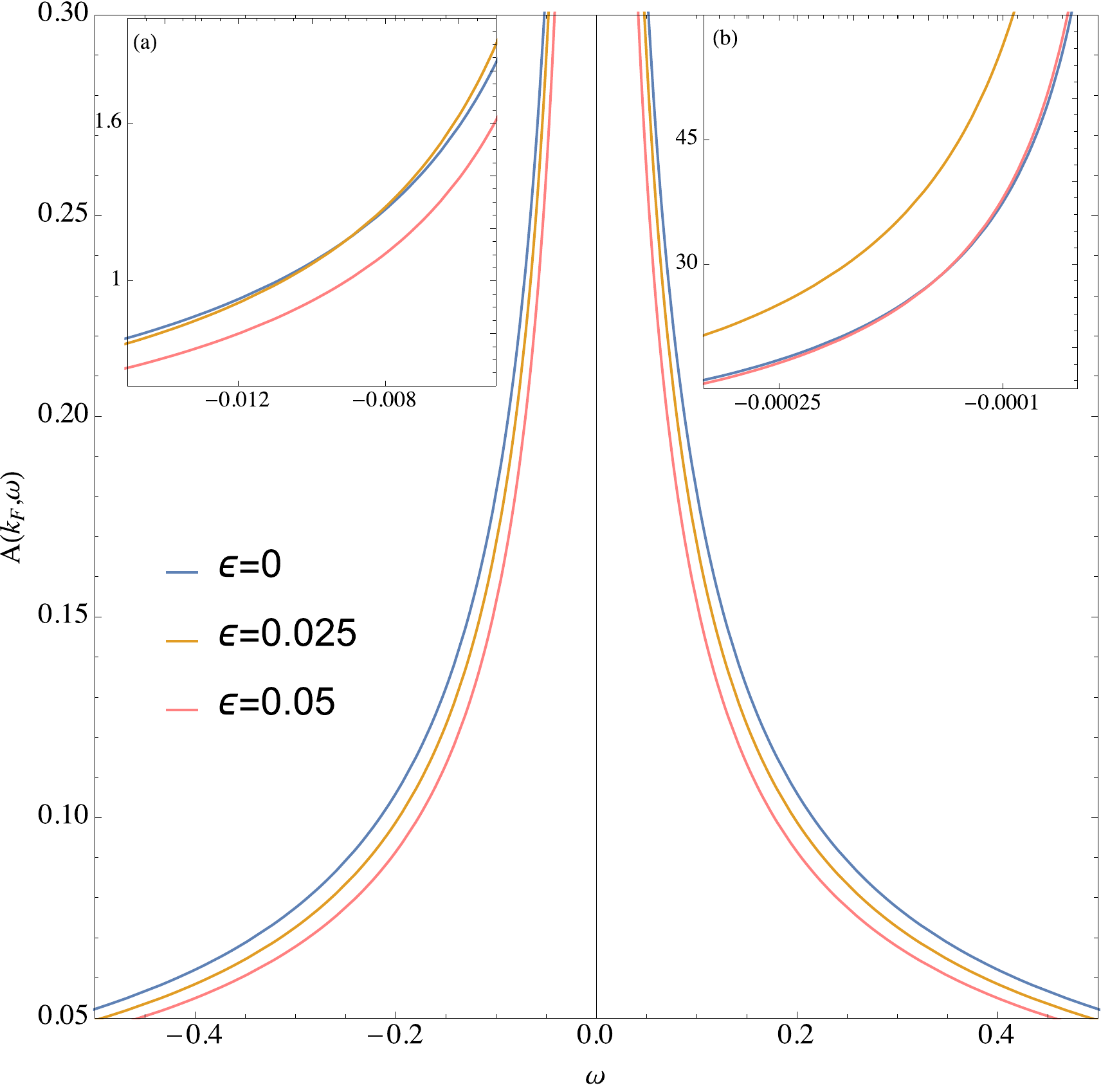,width=0.6\columnwidth}}
  \caption{\label{A} The spectral function at Fermi surface is plotted for $\epsilon=0,\,0.025$ and $0.05$ with $g_{10}=0$, $g_{20}=3\,(\eta=9/(4\pi^2)\approx 0.228)$. The $1+\epsilon$ and 1-d spectral functions intersect when $\epsilon<\eta$, as explained in the text. Inset (a) shows the intersection between spectral functions in  1-d and  ones with $\epsilon=0.025$, and Inset (b) shows the intersection between 1-d and  ones with $\epsilon=0.5$. 
  }
\end{figure}

Next we calculate the spectral function. First we need the real part of the self energy, from which we can compute
\beq
A(k_F,\omega)= -\frac{1}{\pi}  \frac{\Im m \, \Sigma(k_F,\omega)}{(\omega- \Re e \, \Sigma(k_F,\omega))^2+(\Im m \, \Sigma(k_F,\omega))^2}.
\eeq
To the leading order, as in \disp{legf} and \disp{imsigma-leading},
\beq
\Re e[G^{-1}(\omega; \varLambda_{n+1})]=\omega-\Re e \Sigma(k=0,\omega)=[Z (\omega)]^{-1}\omega,
\eeq
so the real part of self-energy at the Fermi surface for small $\omega$ is
\beq
\Re e \Sigma(k_F,\omega)=\Re e \Sigma(k=0,\omega)=(1-[Z(\omega)]^{-1})\omega= (1-\big|\frac{\omega}{2}\big| ^{f(\omega, \epsilon)-1-\epsilon})\omega
\eeq
Therefore the resulting spectral function is $A(k_F,\omega)\propto \omega^{\epsilon-1}$ for small $\omega$, while the 1-d low energy Tomonaga-Luttinger liquid spectral function\cite{Meden,Kopietz2} is $A_{1-d}(k_F,\omega)\propto \omega^{\eta-1}$. We would expect an intersection between the $1+\epsilon$ and the 1-d spectral functions when $\epsilon<\eta$ because $A(k_F,\omega)$ in $1+\epsilon$ dimensions is more singular than that in 1-d. This intersection is shown in \figdisp{A}. The intersection shows the crossover behavior, but one can not distinguish a spectral function as corresponding to $1+\epsilon$ dimensions or 1-d by looking at the low energy behaviors because one can fix $\epsilon$ and change $\eta$ till $\eta=\epsilon$ or vise versa. The important feature of the $1+\epsilon$ dimensional spectral function is that the exponent depends not on the interaction but on the dimension. 

We can also obtain the $k$ dependent self energy and spectral function for small $k$ by noting that the symmetry valid in one dimension that the Greens function, Self energy, etc., depend only on the combination  $\omega-k$ is approximately maintained in $1+\epsilon$ dimensions when $k\to0$, as discussed in more detail in the next section. Hence 
\beq
\begin{split}
\frac{d\Re e \Sigma(k,\omega)}{dk}\Big|_{k\to 0}
&=-\frac{d\Re e \Sigma(k_F,\omega)}{d\omega}
=-1+\frac{1}{Z(\omega)}+\frac{1}{Z(\omega)}\frac{[g(\omega)]^2\big|\frac{\omega}{2}\big|^{\epsilon}\sgn(\omega)}{8\pi^2(1+\epsilon)}
\end{split}
\eeq


\section{Breaking of the $\omega- k $ symmetry in $1+\epsilon$ dimensions\label{symmetrybreaking}}
In one dimension, there is a symmetry by which  the one particle Green function for the right moving electrons close to the Fermi level depends on $\omega$ and $k$ only in the  combination  $\omega-k$. (For the left moving electrons near the fermi level the combination is $\omega+k$.) This symmetry gets broken in higher dimensions. In our previous discussions, we only calculated the case with $k=0$, so the extent of validity or breaking of this symmetry has not been explicitly discussed. We will do so next, and explore to what extent the symmetry is broken in $1+\epsilon$ dimensions by treating the case with $k \neq 0$.

Looking at the $(n+1)^{th}$ step of EPMS, without loss of generality, we choose $k$ and $n$ such that $\varLambda_{n+1}>k>0$. Then we have (compare with Eq.s \ref{Ig2} and \ref{III})
\beq
\begin{split}
I(k,i\omega;\varLambda_n)&=[g(\varLambda_n)]^2s^n\int_{-\varLambda_n}^{\varLambda_n}\frac{|k_1|^{\epsilon} dk_1}{2\pi}\int_{-\varLambda_n}^{\varLambda_n} \frac{dk_2}{2\pi}\int_{-\infty}^{+\infty}\frac{d\omega_1}{2\pi}\int_{-\infty}^{+\infty}\frac{d\omega_2}{2\pi}
\\&\frac{1}{i\omega_1+k_1}\frac{1}{i\omega_2-k_2}\frac{1}{i(\omega_1+\omega_2-\omega)+(k_1+k_2-k)}.\label{I-kw-in-1+eps}
\end{split}
\eeq
Performing the frequency integrals using contour integration as before, it is not hard to see that one set of non-vanishing contributions to $I(k, i\omega;\varLambda_n)$ arise when $(-\varLambda_n<k_1<0$, $-\varLambda_n<k_2<0$, $-\varLambda_n<k_1+k_2-k<0)$. After simplifying, this corresponds to the conditions that either $(0<k_1<k-k_2$ and $k-\varLambda_n<k_2<0)$ or $(-\varLambda_n+k-k_2<k_1<\varLambda_n$ and $-\varLambda_n<k_2<k-\varLambda_n)$. Hence
we get the contributions (compare Eq. \ref{I1} )
\beq
\begin{split}
\frac{I_1(k, i\omega; \varLambda_n)}{s^n[g(\varLambda_n)]^2}
&=\int^{\varLambda_n-k}_0\frac{dk_2}{2\pi}\int_0^{k_2+k}\frac{|k_1|^\epsilon dk_1}{2\pi}\frac{1}{i\omega+2k_2+k}+\int^{\varLambda_n}_{\varLambda_n-k}\frac{dk_2}{2\pi} \int_{-\varLambda_n+k+k_2}^{\varLambda_n}\frac{|k_1|^\epsilon dk_1}{2\pi}\frac{1}{i\omega+2k_2+k}
\end{split}\label{I1-k}
\eeq
The other set of nonvanishing contributions to $I(k, i\omega;\varLambda_n)$ arise when $(0<k_1<\varLambda_n$, $\varLambda_n>k_2>0$, $\varLambda_n>k_1+k_2-k>0$), which, after simplifying, leads to the conditions  $-k_2+k<k_1<0$ and $k<k_2<\varLambda_n$. So we get the second contribution to be (compare Eq. \ref{I2})
\beq
\begin{split}
\frac{I_2(k, i\omega; \varLambda_n)}{s^n[g(\varLambda_n)]^2}
=\int^{\varLambda_n}_k\frac{dk_2}{2\pi}\int_0^{k_2-k} \frac{|k_1|^\epsilon dk_1}{2\pi}\frac{1}{i\omega-2k_2+k}.
\end{split}\label{I2-k}
\eeq
Assuming $k\ll\varLambda_{n}$, we can neglect the second term in $I_1$. Then, after making the analytic continuation $i\omega \to \omega^+$, we get, for the incremental contribution to the $k$,$\omega$ dependent self energy as per the EPMS prescription,
\beq
\begin{split}
\frac{\Delta I(k, \omega^+; \varLambda_n, \varLambda_{n+1})}{s^n[g(\varLambda_n)]^2}& \equiv \frac{I_1(k, \omega^+; \varLambda_n)+I_2(k, \omega^+; \varLambda_n)}{s^{n}[g(\varLambda_{n})]^2} - \frac{I_1(k, \omega^+; \varLambda_{n+1})+I_2(k, \omega^+; \varLambda_{n+1})}{s^{n+1}[g(\varLambda_{n+1})]^2}
\\&=\int^{\varLambda_n-k}_{\varLambda_{n+1}-k}\frac{|k_2+k|^\epsilon dk_2}{4\pi^2(1+\epsilon)}\frac{1}{2} (1-\frac{\omega^+ -k}{\omega^+ +2k_2+k})+\int^{\varLambda_n}_{\varLambda_{n+1}}\frac{|k_2-k|^\epsilon dk_2}{4\pi^2(1+\epsilon)}\frac{1}{2}(-1+\frac{\omega^+ -k}{\omega^+ -2k_2+k}).
\\&\approx \int^{\varLambda_n}_{\varLambda_{n+1}}\frac{|k_2|^\epsilon dk_2}{4\pi^2(1+\epsilon)}\frac{1}{2} (1-\frac{\omega^+ -k}{2k_2})+\int^{\varLambda_n}_{\varLambda_{n+1}}\frac{|k_2|^\epsilon dk_2}{4\pi^2(1+\epsilon)}\frac{1}{2}(-1+\frac{\omega^+ -k}{-2k_2})
\\&=-\int^{\varLambda_n}_{\varLambda_{n+1}}\frac{|k_2|^\epsilon dk_2}{4\pi^2(1+\epsilon)}\frac{\omega^+ -k}{2k_2}
\end{split}\label{delI-kw}
\eeq
where we have retained only leading term ($\omega, k \ll \varLambda_{n+1}$) in the last two steps. When $k\rightarrow 0$, we go back to the leading term of \disp{deltaI}. We see that the extra $\epsilon$ dimension brings an factor which only depends on $k$ and therefore breaks the symmetry. But for a small enough  $k$, the symmetry can still be thought of as approximately maintained.

\section{Conclusions\label{conclusion}}

In summary, we have presented the low energy Greens function of interacting fermions in $1+\epsilon$ dimensions,
with the momentum fixed at the Fermi point. Going beyond the  lowest order terms  in literature, our calculation  gives  the leading behavior of the damping of the quasiparticles. In order to obtain this result we  extended the poor man's scaling method of Anderson into a Wilsonian type framework. The extended poor man's scaling method developed and used here retains the appealing features of Anderson's scaling and is applied to the Green's functions rather than the Hamiltonian itself as in the original method, and we expect that  this might be useful in other contexts as well.

Our work shows  that the  Tomonaga-Luttinger  behavior seen in 1-dimension  is destroyed  for nonzero $\epsilon$, and the system becomes a Fermi liquid with a small but finite $Z=\exp(-d_0\,\eta^{3/4}/\sqrt{\epsilon})$, so that for $\epsilon<\eta$ the  $Z$ decreases non analytically with $\epsilon$. Further at low $\omega$ the damping rate is calculated and found to be $\sim |\omega|^{1+\epsilon}/{\log(|\omega|/2)^2}$. Thus the damping  is smaller than the particle energy $\omega $ and supports the notion of a ``Fragile Fermi Liquid'', one where the quasiparticles are rendered fragile by the small magnitude of $Z$. The damping rate found here, while  different from the familiar $\omega^2$ behavior of a  3-d Fermi liquid, exhibits  an $\omega/Z$ scaling, apart from a weak logarithmic correction term. This results in an explicit
 low frequency i.e. quasiparticle Green function given in \disp{gsimple3}.  {Although \disp{result-damping-1}, (\ref{result-damping-2}) and (\ref{gsimple3}) depend on our choice of the specific $1+\epsilon$ prescription, we expect only quantitative differences would result from other $1+\epsilon$ schemes.}

While this fragile FL behavior is seen at the lowest $\omega$, we find a crossover to a Tomonaga-Luttinger behavior at higher $\omega$.  The crossover behavior is captured  in the flow equations of coupling constants and quasiparticle weight as well as damping term with a crossover scale $l^*=1/\epsilon$ or $\omega^*=2\textrm{e}^{1/\epsilon}$. When $l\ll l^*$ or $\omega\gg\omega^*$, the system behaves like a Tomonaga-Luttinger liquid, while in the other limit $l\gg l^*$ or $\omega\ll\omega^*$, it displays a fragile Fermi liquid behavior. 

We also computed the electron spectral function for typical values of parameters in $1+\epsilon$ dimensions in \figdisp{A}. This result could cast some light on the expected spectral functions in    ARPES experiments on coupled linear chain compounds. A cautionary remark is due here, since the spectral function for  both the purely 1-d and $1+\epsilon$ dimensional cases diverge as power laws at $\omega\to 0$, distinguishing between them from such plots  is not an easy task. These results might be helpful  in designing further experiments to test the theory quantitatively.

\section{Acknowledgements}
The work at University of California, Santa Cruz (UCSC) was supported by the U.S. Department of Energy (DOE), Office of Science, Basic Energy Sciences (BES) under Award \# FG02-06ER46319. HRK would also like to acknowledge support from the Department of Science and Technology, India. We thank G-H Gweon,W. Metzner,  S. Raghu, R. Shankar for helpful comments and discussions.

\appendices
\section{ Wilsonian RG in interacting fermion systems \label{Shankar-RG}}
There are three steps in the Wilsonian RG for fermions, as explained in Shankar~\cite{Shankar}. The first step is mode elimination. Let us express the partition function as:
\beq
{\cal Z}=\int[{\cal D}\phi_<][{\cal D}\phi_{>}]\textrm{e}^{S_0(\phi_{<})}\textrm{e}^{S_0(\phi_{>})} \textrm{e}^{S_I(\phi_{<},\phi_{>})}
\eeq
where $\phi_<$ and $\phi_>$ denote $\phi(k)$ for $0\leq |k| \leq \varLambda_0/s$ (slow modes) and $\varLambda_0/s \leq |k| \leq \varLambda_0$ (fast modes) respectively, $S_0$ is the quadratic part of the action and $S_I$ is the interaction part. The effective action $S'(\phi_{<})$ is defined such that
\beq
{\cal Z}=\int[{\cal D}\phi_<]\textrm{e}^{S'(\phi_{<})}
\eeq
Clearly,
\beq
\begin{split}
\textrm{e}^{S'(\phi_{<})}&=\textrm{e}^{S_0(\phi_{<})}\int[{\cal D}\phi_{>}]\textrm{e}^{S_0(\phi_{>})} \textrm{e}^{S_I(\phi_{<}, \phi_{>})}
=\textrm{e}^{S_0(\phi_{<})}\langle \textrm{e}^{S_I(\phi_{<}, \phi_{>})} \rangle_{0>}
\end{split}
\eeq
where $\langle\rangle_{0>}$ stands for averages with respect to the fast modes, and  $\int[{\cal D}\phi_{>}]\exp[S_0(\phi_{>})]$, a constant which will not affect any correlation functions of slow modes, has been dropped. After mode eliminations, there are two more steps. Suppose we had an initial action:
\beq
\begin{split}
S(\phi)&=\int^\infty_{-\infty} \frac{d\omega}{2\pi}\int^{\varLambda_0}_{-\varLambda_0}\frac{dk}{2\pi}  (i\thinspace\omega-k) \phi^{*}(k\thinspace\omega)\phi(k\thinspace\omega)+\int_{k\thinspace\omega; \varLambda_0}u_2(1, 2, 3, 4)\phi^{*}(1)\phi^{*}(2)\phi(3) \phi(4)\delta(1+2-3-4)
\end{split}
\eeq
After the fast modes are integrated out, the momentum cut-off in the effective action reduces to $\varLambda_0/s$. We can write the effective action in the form
\beq
\begin{split}
S'(\phi_<)&=\int^\infty_{-\infty} \frac{d\omega}{2\pi}\int^{\varLambda_0/s}_{-\varLambda_0/s}\frac{dk}{2\pi}  [a\thinspace i\thinspace\omega-b \thinspace k+o(\omega^2, k^2, k\omega)] \phi^{*}_<(k\thinspace\omega)\phi_<(k\thinspace\omega)
\\&+\int_{k\thinspace\omega; \varLambda_0/s}c\thinspace u_2(1, 2, 3, 4)~\phi^{*}_<(1)\phi^{*}_<(2)\phi_<(3) \phi_<(4) ~\delta(1+2-3-4)
\\&+\int_{k\thinspace\omega; \varLambda_0/s}u_3(1, 2, 3, 4, 5, 6)... + ....~,\label{Sphiless}
\end{split}
\eeq
where $a$, $b$ and $c$ are \textit{multiplicative}  renormalization coefficients of the various couplings arising from mode elimination. There are also some higher order terms like the 3 particle interaction term  $u_3$ (not written out in detail), 4 particle interaction terms, etc., that result from the mode elimination. It is useful  to have an effective action that looks as much like as our original action as possible~\cite{Wilsonkogut}. So we define new momenta, $k'=sk$, and new frequencies, $\omega'=s\omega$ to represent the effective action.  And we also rescale the fields as 
\beq
\phi'(k',\omega')=\zeta^{-1}\phi_<(k, \omega).\label{fscaling}
\eeq
where conventionally~\cite{Shankar} $\zeta=s^{3/2}/a^{1/2}$ is chosen to fix the coefficient of $i\omega'$ in the quadratic part of action to stay as 1. So the effective action can be written as
\beq
\begin{split}
S'(\phi')&=\int^\infty_{-\infty} \frac{d\omega'}{2\pi}\int^{\varLambda_0}_{-\varLambda_0}\frac{dk'}{2\pi}   (i\omega'-\frac{b}{a}k'+o(\omega'^2/s, k'^2/s, k'\omega'/s)) \phi'^{*}(k'\thinspace\omega')\phi'(k'\thinspace\omega')
\\&+\int_{k'\thinspace\omega';\varLambda_0}u_2'(1', 2', 3', 4')\phi'^{*}(1')\phi'^{*}(2')\phi'(3') \phi'(4')
\delta(1'+2'-3'-4')\\&+ \int_{k'\thinspace\omega';\varLambda_0} u'_3(1', 2', 3', 4', 5' , 6') ... + ...  ~,\label{Seff}
\end{split}
\eeq
Here the cutoff for $k'$ has been restored to the  original cutoff $\varLambda_0$  and $\delta(1'+2'-3'-4')=\delta(k'_1+k'_2-k'_3-k'_4)\delta(\omega'_1+\omega'_2-\omega'_3-\omega'_4)$; $u_2'(1', 2', 3', 4')$ is a shorthand notation for $u_2'(k_1'\thinspace\omega_1', k_2'\thinspace\omega_2', k_3'\thinspace\omega_3', k_4'\thinspace\omega_4')$, and likewise for $u'_3(1', 2', 3', 4', 5', 6')$, etc. Invoking the definition of renormalized couplings, we get\textit{}
\beq
u_2'(1', 2', 3', 4')=\frac{\zeta^4 c u_2(1, 2, 3, 4)}{s^{6}} = \frac{cu_2(1, 2, 3, 4)}{a^2},\label{twobody}
\eeq
\beq
u'_3(1', 2', 3', 4', 5', 6')=\frac{\zeta^6  u_3(1, 2, 3, 4, 5, 6)}{s^{10}}=\frac{u_3(1, 2, 3, 4, 5, 6)}{sa^3}.\label{threebody}
\eeq
Expanding the couplings in powers of the momenta and the frequencies, we can see from Eqs. (\ref{Seff}) and (\ref{twobody}) that the leading one body terms and the two body couplings that are independent of the momenta and the frequencies  are marginal in the RG sense, because the power of $s$ involved in their linear recursion relations is $0$;  while the three body coupling is irrelevant due to the negative power of $s$ in \disp{threebody}, as is also the case for the non leading one body terms, and the frequency and momentum dependent two body terms. After $n$ such steps of the RG, we take the infinitesimal mode elimination limit $s=\textrm{e}^{dl}$ and $dl\rightarrow0$,   holding  $n \, dl=l$ fixed, whence the running cutoff becomes $\varLambda_n \rightarrow \varLambda=\varLambda_0 \textrm{e}^{-l}$. Then we get the differential equations (RG flow equations) for the cutoff dependent quantities like the coupling constants, examples of which are presented in Section \ref{g1g2} of the text.

Next, we discuss the changes in the Green function during the RG process.   After the first step of the mode elimination, we can clearly  calculate    low energy ($|k|,|\omega|\ll \varLambda_1$) Green functions of the \textit{original} fields using the effective action with the reduced cutoff $\varLambda_1$ (cf., Eq.~\ref{Sphiless}):
\beq
G(k, i\omega;\varLambda_1){\delta(k-q)\delta(\omega-\nu) }\equiv \langle \phi_< (k\thinspace\omega) \phi^{*}_< (q\thinspace \nu) \rangle_{S'(\phi_<)}~.
\eeq
The corresponding  Green function involving the \textit{rescaled} fields  is (cf., Eq.~\ref{Seff})
\beq
G_1(k', i\omega';\varLambda_0)\delta(k'-q')\delta(\omega'-\nu')\equiv  \langle\phi'(k'\thinspace\omega') \phi'^{*}(q'\thinspace\nu') \rangle_{S'(\phi')}~.
\eeq
The momentum cutoff arguments in the Green functions refer to the cutoffs in  the effective actions using which the Green functions are being evaluated. From the relations between the original and the rescaled fields (cf., Eq. \ref{fscaling}), it is straightforward to verify that the  two Green functions are related  as
\beq
G(k, i\omega;\varLambda_1)= s^{-2}\zeta^2 G_1(k', i\omega';\varLambda_0) =a^{-1}\thinspace s\thinspace G_1(k', i\omega';\varLambda_0), \label{gfscaling}
\eeq
with $k'=sk$ and $\omega'=s\omega$.

Consider doing perturbative calculations of these. Keeping only the leading order terms in $k$ and $\omega$,  to \textit{zeroth order} in the couplings we get (cf., Eq.~\ref{Sphiless})
\beq
G^{(0)}(k, i\omega;\varLambda_1)=\frac{1}{a\thinspace i\omega-b\thinspace k}~,
\eeq
whereas the corresponding lowest order Green function involving the \textit{rescaled} fields  is (cf., Eq.~\ref{Seff})
\beq
G_1^{(0)}(k', i\omega';\varLambda_0) =\frac{1}{i\omega'-(b/a)k'}~,
\eeq
with the relationship between the two in agreement with \disp{gfscaling}.

More generally, after $n$ steps of the RG reducing the cutoff to $\varLambda_n \equiv \varLambda_0/s^n$, the corresponding relation between the two Green functions is
\beq
G(k, i\omega;\varLambda_n)=s^{-2n}[\zeta(\varLambda_n)]^2G_n(k_n, i\omega_n;\varLambda_0) =[a(\varLambda_n)]^{-1}s^n G_n(k_n, i\omega_n;\varLambda_0),\label{gfrelation}
\eeq
with $\omega_n \equiv \omega s^n$, $k_n \equiv k s^n$,   $a(\varLambda_n)$ and $b(\varLambda_n)$ being the \textit{accumulated} multiplicative renormalization coefficients, and $\zeta(\varLambda_n)$  being the corresponding field rescaling factor.
 As can be seen for example by iterating the relation \ref{gfscaling}, the recursive nature of RG allows us to write
\beq
a(\varLambda_n)=\prod_{m=0}^{n-1}\tilde{a}(\varLambda_m \rightarrow \varLambda_{m+1}).\label{arelation}
\eeq
\beq
b(\varLambda_n)=\prod_{m=0}^{n-1}\tilde{b}(\varLambda_m \rightarrow \varLambda_{m+1})\label{brelation}
\eeq
Here  $\tilde{a}(\varLambda_m \rightarrow \varLambda_{m+1})$ and $\tilde{b}(\varLambda_m \rightarrow \varLambda_{m+1})$ are the (multiplicatively) incremental renormalization factors for the coefficients of the frequency and of the momentum respectively, due to the RG step that reduces the running cutoff from $\varLambda_m$ to $\varLambda_{m+1}$. For example, $\tilde{a}(\varLambda_0 \rightarrow \varLambda_1)$ and $\tilde{b}(\varLambda_0 \rightarrow \varLambda_1)$ correspond to the $a$ and $b$ in \disp{Sphiless}. Needless to say, $a(\varLambda_0)=1$ and $b(\varLambda_0)=1$. The relation in \disp{gfrelation} is consistent with that in~\refdisp{Shankar,Wilsonkogut}. \beq
G_n^{(0)}(k_n,i\omega_n; \varLambda_0)=\frac{1}{i\omega_n-[b(\varLambda_n)/a(\varLambda_n)]k_n}
\eeq
is the lowest order Green function of the rescaled fields after $n$ steps of RG.

\section{Detailed discussion of the Extended Poor Man's Scaling method\label{fulldetail}}

In this appendix, we show explicitly that  the second order EPMS prescription presented in the text includes the same second order diagrams as second order Wilsonian RG does if we reduce the running cutoff from $\varLambda_0$ to $0$. Specifically, we show that  the total contribution from formally irrelevant two and three body vertexes arising at  a certain step of mode elimination in Wilsonian RG to the self energy in all the subsequent steps of the RG is equal to the result from that same mode elimination in EPMS.
\begin{figure}
  \centerline{\epsfig{file=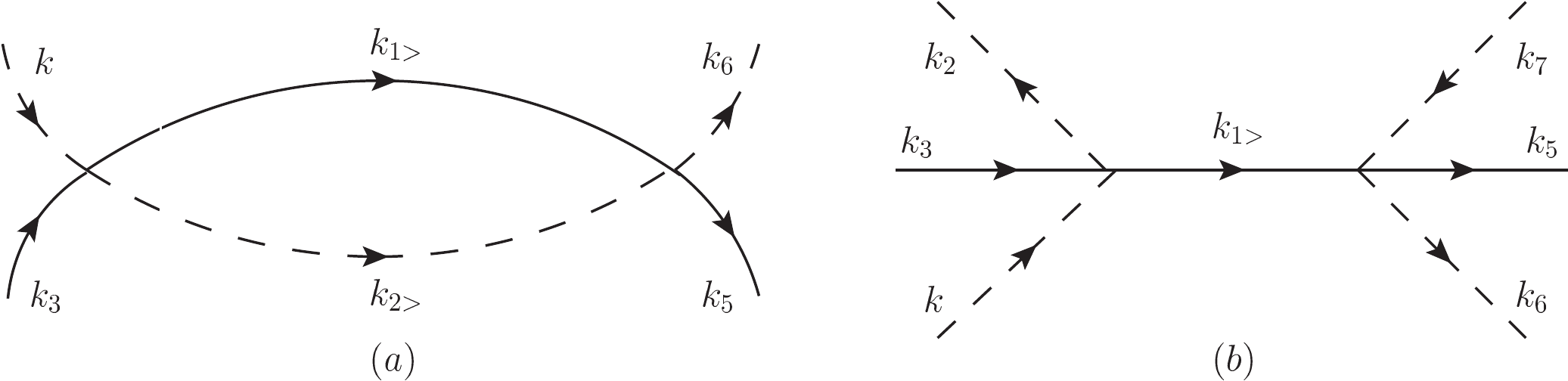,width=0.5\columnwidth}}
  \caption{\label{23ver} (a) is an example for the two body vertex and (b) is a three body vertex.
  }
\end{figure}
We focus on the renormalization of the one body vertex, which is the same as the self energy, with $k$ chosen on the right branch without loss of generality. Consider the three body vertex produced at the $(m_1+1)^{th}$ step of RG, as shown in \figdisp{23ver} (b). There are two ways for this vertex to contribute to the self energy in later steps of the RG. One is to have it first renormalize the two body vertex by integrating out a second momentum (say $k_2$) at the $(m_2+1)^{th}$ step of RG, and then have it renormalize the one body vertex by integrating out the third momentum (say $k_3$) at the $(m_3+1)^{th}$ step of RG; and vice versa. The other way is to have it renormalize the one body vertex at one shot by integrating out the second and third momenta at the $(m_2+1)^{th}$ step of RG. We look at the former way first, since it is more general, and the second way can be obtained from it simply by setting $m_3=m_2$.

The three body vertex (in \figdisp{23ver} b) produced from mode elimination in the $(m_1+1)^{th}$ RG step is
\beq
\begin{split}
&u_{3m_1}(0_{m_1}, 2_{m_1}, 3_{m_1}, 5_{m_1}, 6_{m_1}, 7_{m_1})
\\&=[g(\varLambda_{m_1})]^2\int_{d\varLambda_0}\frac{dk_{1m_1}}{2\pi}\int_{-\infty}^{\infty}\frac{d\omega_{1m_1}}{2\pi}\frac{1}{i\omega_{1m_1}+k_{1m_1}}\delta(1_{m_1}+2_{m_1}-3_{m_1}-0_{m_1})\label{three},
\end{split}
\eeq
where $u_{3m_1}(0_{m_1}, 2_{m_1}, 3_{m_1}, 5_{m_1}, 6_{m_1}, 7_{m_1}) \equiv u_{3m_1}(k_{m_1}\thinspace i\omega_{m_1}, k_{2m_1}\thinspace i\omega_{2m_1}, k_{3m_1}\thinspace i\omega_{3m_1}, k_{5m_1}\thinspace i\omega_{5m_1}, k_{6m_1}\thinspace i\omega_{6m_1}, k_{7m_1}\thinspace i\omega_{7m_1})$, $\delta(1_{m_1}+2_{m_1}-3_{m_1}-0_{m_1})\equiv \delta(k_{1m_1}+k_{2m_1}-k_{3m_1}-k_{m_1})\delta(\omega_{1m_1}+\omega_{2m_1}-\omega_{3m_1}-\omega_{m_1})
$, $\omega_{am}=\omega_{a}s^{m}$ and $k_{am}=k_{a}s^{m}$. 
There are other three body interaction terms arising from previous steps of the RG, but we focus on this specific term appearing due to mode elimination at the $(m_1+1)^{th}$ RG step and see how it contributes to the two body and one body vertexes in subsequent steps of the RG.
Next, at the $(m_2+1)^{th}$step of the RG, this three body interaction produces a two body coupling (in \figdisp{23ver} a) to linear order upon integration over $k_2$ and $\omega_2$: 
\beq
\begin{split}
&u_{2m_2}(0_{m_2}, 3_{m_2}, 5_{m_2}, 6_{m_2})=\frac{1}{s^{m_2-m_1}}\int_{d\varLambda_0}\frac{dk_{2m_2}} {2\pi}\int^\infty_{-\infty}\frac{d\omega_{2m_2}}{2\pi}\frac{1}{i\omega_{2m_2}-k_{2m_2}}u_{3m_1}(0_{m_1}, 2_{m_1}, 3_{m_1}, 5_{m_1}, 6_{m_1}, 2_{m_1}),\label{two}
\end{split}
\eeq
where the factor $1/s^{m_2-m_1}$, arising from a combination of the scaling of the momenta and of the fields as discussed earlier, ensures the formal irrelevance of the three body vertex $u_3$ as in \disp{threebody}. Here we have neglected the corrections arising from the $1/a$ factors in \disp{threebody} because we are only looking at the second order contributions, while $1/a=1+o(g^2)$ would introduce higher order contributions. 
Then at the $(m_3+1)^{th}$ step of the RG, again to linear order in the interaction, this two body interaction produces a one body coupling upon integration
\beq
\begin{split}
u_{1m_3}(0_{m_3})=\int_{d\varLambda_0}\frac{dk_{3m_3}} {2\pi}\int^\infty_{-\infty}\frac{d\omega_{3m_3}}{2\pi}\frac{1}{i\omega_{3m_3}-k_{3m_3}}u_{2m_2}(0_{m_2}, 3_{m_2}, 3_{m_2}, 0_{m_2})\label{one},
\end{split}
\eeq
which renormalizes the original one body vertex. We can do these three integrations in one line, denoting the resulting one body vertex as $P_{m_1m_2m_3}(k, i\omega)$:
\beq
\begin{split}
P_{m_1m_2m_3}(k, i\omega)
=[g(\varLambda_{m_1})]^2s^{m_3}(\prod_{j=1}^3\int_{d\varLambda_{m_j}} \frac{dk_j}{2\pi}\int^\infty_{-\infty}\frac{d\omega_j}{2\pi})\frac{1}{i\omega_1+k_1}\frac{1}{i\omega_2-k_2}\frac{1}{i\omega_3+k_3}
\delta(1+2-3-0)\label{P123}
\end{split}
\eeq
where the momenta and frequencies have been restored to their original scales, whence the momentum integrals are over the appropriate momentum shells, with $d\varLambda_{m_j}$ denoting the momentum shell $\varLambda_{m_j+1}<|k_j|<\varLambda_{m_j}$. The reason why $m_1$, $m_2$, $m_3$ are not the same is that these integrals are not done at the same step of the RG.

As mentioned above, the second type of contribution to the one-body term can also be included in this formalism simply as a special case, with $m_2=m_3>m_1$. And so can the (vanishing) sunrise diagram in \figdisp{sunrise} (a), by setting  $m_1=m_2=m_3$. In fact, all contributions to the one-body term coming from the two or three body vertex produced in any arbitrary [say $(n+1)^{th}$] step of Wilsonian's RG can be represented in terms of $P_{m_1m_2m_3}(k,\omega)$. From \disp{P123}, it is not hard to see that we can \textit{redefine} $P_{m_1m_2m_3}(k, i\omega)$ for an \textit{arbitrary} choice of $m_1$, $m_2$ and $m_3$ as,
\beq
\begin{split}
P_{m_1m_2m_3}(k, i\omega)=&[g(\varLambda_{m_s})]^2s^{m_l}(\prod_{j=1}^3\int_{d\varLambda_{m_j}} \frac{dk_j}{2\pi}\int^\infty_{-\infty}\frac{d\omega_j}{2\pi})\frac{1}{i\omega_1+k_1}\frac{1}{i\omega_2-k_2} \frac{1}{i\omega_3+k_3}\delta(1+2-3-0)\label{P123g},
\end{split}
\eeq
where $m_l \equiv \max(m_1, m_2, m_3)$ and $m_s \equiv \min(m_1,m_2,m_3)$. Then we can calculate the self-energy term arising from the the two or three body vertexes produced at the $(n+1)^{th}$ step of the RG  as the constrained sum:
\beq
\begin{split}
\Delta Q(k, i\omega; \varLambda_n, \varLambda_{n+1})&=\sum_{m_1, m_2,m_3}\delta_{m_l n}~P_{m_1,m_2,m_3}(k, \omega)
\approx Q(k, i\omega; \varLambda_{n+1})/s-Q(k, i\omega; \varLambda_n),
\end{split}
\eeq
where
\beq
\begin{split}
Q(k, i\omega; \varLambda_n)& \equiv s^{n}(\prod_{j=1}^3\int_{\varLambda_n}^{\varLambda_0} \frac{dk_j}{2\pi}\int^\infty_{-\infty}\frac{d\omega_j}{2\pi}+\prod_{j=1}^3\int^{-\varLambda_n}_{-\varLambda_0} \frac{dk_j}{2\pi}\int^\infty_{-\infty}\frac{d\omega_j}{2\pi})[g(|k_l|)]^2\frac{1}{i\omega_1+k_1}\frac{1}{i\omega_2-k_2}\frac{1}{i\omega_3+k_3}
\\&\delta(1+2-3-0)\end{split}
\eeq
with  $|k_l| \equiv \max(|k_1|, |k_2|, |k_3|)$. The approximation above  becomes exact when we take the infinitesimal mode elimination limit. Hence, for the incremental contributions to the multiplicative renormalization coefficients,
we obtain\beq
\tilde{a}_w(\varLambda_m \rightarrow \varLambda_{m+1})=1+\frac{\partial \Delta Q(k, i\omega; \varLambda_m, \varLambda_{m+1})}{\partial (i\omega_m)}\Big|_{\omega\rightarrow 0,k\rightarrow 0}
\eeq
and
\beq
\tilde{b}_w(\varLambda_m \rightarrow \varLambda_{m+1})=1+\frac{\partial \Delta Q(k, i\omega; \varLambda_m, \varLambda_{m+1})}{\partial k_m}\Big|_{\omega\rightarrow 0,k\rightarrow 0}.
\eeq
The subscript "$w$" is for reminding ourselves that these contributions are from the Wilsonian RG. Next, we can compare $a_w(\varLambda_n)$ and $a_e(\varLambda_n)$ for example. By definition,
\beq
a_w(\varLambda_n)=\prod_{m=0}^{n-1}\tilde{a}_w(\varLambda_m \rightarrow \varLambda_{m+1})=\prod_{m=0}^{n-1} (1+\frac{\partial \Delta Q(k, i\omega; \varLambda_m, \varLambda_{m+1})}{\partial (i\omega_m)}\Big|_{\omega\rightarrow 0,k\rightarrow 0})\label{ca1}
\eeq

Using Eqs. (\ref{III})-(\ref{b2}), we get
\beq
a_e(\varLambda_n)=\prod_{m=0}^{n-1}\tilde{a}_e(\varLambda_m \rightarrow \varLambda_{m+1})=\prod_{m=0}^{n-1}(1+\frac{\partial \Delta I(k, i\omega; \varLambda_m, \varLambda_{m+1})}{\partial (i\omega_m)}\Big|_{\omega\rightarrow 0,k\rightarrow 0})\label{ca2}
\eeq
It is not hard to verify that
\beq
\begin{split}
\sum^\infty_{m=0}\frac{\Delta I(k, i\omega; \varLambda_m, \varLambda_{m+1})}{s^m}&\approx\prod_{j=1}^3\int_{-\varLambda_0}^{\varLambda_0} \frac{dk_j}{2\pi}\int^\infty_{-\infty}\frac{d\omega_j}{2\pi}[g(|k_l|)]^2\frac{1}{i\omega_1+k_1}\frac{1}{i\omega_2-k_2}\frac{1}{i\omega_3+k_3}\delta(1+2-3-0)
\\&=\lim_{n\rightarrow\infty}\frac{Q(k, i\omega; \varLambda_n)}{s^n}=\sum^\infty_{m=0}\frac{\Delta Q(k, i\omega; \varLambda_m, \varLambda_{m+1})}{s^m}.
\end{split}
\eeq
where the approximation again becomes exact when we take the infinitesimal mode elimination limit, and $\varLambda_{\infty}=0$. So,
\beq
\begin{split}
\sum^\infty_{m=0}\frac{\partial \Delta I(k, i\omega; \varLambda_m, \varLambda_{m+1})}{\partial (i\omega_m)}\Big|_{\omega\rightarrow 0,k\rightarrow 0}
=\sum^\infty_{m=0}\frac{\partial \Delta Q(k, i\omega; \varLambda_m, \varLambda_{m+1})}{\partial (i\omega_m)}\Big|_{\omega\rightarrow 0,k\rightarrow 0}\label{secor}
\end{split}
\eeq
Set $x_{1m}=[\partial \Delta Q(k, i\omega, \varLambda_m, \varLambda_{m+1})/(\partial (i\omega_m))]\Big|_{\omega\rightarrow 0,k\rightarrow 0}$ and $x_{2m}=[\partial \Delta I(k, i\omega, \varLambda_m, \varLambda_{m+1})/(\partial (i\omega_m))]\Big|_{\omega\rightarrow 0,k\rightarrow 0}$ for convenience. Then, Eqs. (\ref{ca1}), (\ref{ca2}) and (\ref{secor}) can be rewritten as
\begin{eqnarray}
a_w(\varLambda_\infty)=\prod_{m=0}^{\infty}(1+x_{1m}),
\nonumber\\
a_e(\varLambda_\infty)=\prod_{m=0}^{\infty}(1+x_{2m}),
\nonumber\\
\sum_{m=0}^{\infty}x_{1m}=\sum_{m=0}^{\infty}x_{2m}.
\end{eqnarray}
By definition, $x_{1m}\neq x_{2m}$ , but both are small quantities that are of  second order in  the running coupling constants. Hence $a_1(\varLambda_\infty)$ and $a_2(\varLambda_\infty)$ include the same second order Feynman diagrams, but different higher order diagrams, and likewise for $b_1(\varLambda_\infty)$ and $b_2(\varLambda_\infty)$. So we see that when the running cutoff is reduced to $0$, calculations of the self energy using Wilsonian RG and EPMS up to a specific  order in the running coupling constants sum over all the same diagrams up to that order, but in principle different subsets of higher order diagrams.
And as we have argued in this paper, the latter is much easier to implement.


\end{document}